\documentclass[journal,draftcls,onecolumn,12pt]{IEEEtran}
\ifCLASSINFOpdf
\else
\fi

\usepackage{graphicx}
\usepackage{bm}
\usepackage{amsfonts}
\usepackage{amsmath}
\usepackage{amssymb}
\usepackage{times}
\usepackage{subfigure}
\usepackage{latexsym,bm,amsmath,amssymb} 
\usepackage{CJK}
\usepackage{hhline}
\usepackage{cite}

\usepackage{multirow} 
\usepackage{amsmath}
\usepackage{xcolor}
\usepackage[linesnumbered,ruled,vlined]{algorithm2e}
\usepackage{epstopdf}

\begin{document}
%
\title{Performance Analysis of Coded OTFS Systems over High-Mobility Channels}

\author{Shuangyang Li,~\IEEEmembership{Student Member,~IEEE,} Jinhong Yuan,~\IEEEmembership{Fellow,~IEEE,} Weijie Yuan,~\IEEEmembership{Member,~IEEE,} Zhiqiang Wei,~\IEEEmembership{Member,~IEEE,}
Baoming Bai,~\IEEEmembership{Senior Member,~IEEE,} and Derrick Wing Kwan Ng,~\IEEEmembership{Senior Member,~IEEE}
\thanks{
   Part of the paper has been submitted to the IEEE International Conference on Communication Workshops 2021 \cite{Li2020on}.
  }
  \vspace{-5mm}
}

\maketitle

\begin{abstract}
Orthogonal time frequency space (OTFS) modulation is a recently developed multi-carrier multi-slot transmission scheme for wireless communications in high-mobility environments.
In this paper, the error performance of coded OTFS modulation over high-mobility channels is investigated.
We start from the study of conditional pairwise-error probability (PEP) of the OTFS scheme, based on which its performance upper bound of the coded OTFS system is derived. Then, we show that the coding improvement for OTFS systems depends on the squared Euclidean distance among codeword pairs and the number of independent resolvable paths
of the channel.
More importantly, we show that there exists a fundamental trade-off between the coding gain and the diversity gain for OTFS systems, i.e., the diversity gain of OTFS systems improves with the number of resolvable paths, while the coding gain declines.
Furthermore, based on our analysis, the impact of channel coding parameters on the performance of the coded OTFS systems is unveiled.
The error performance of various coded OTFS systems over high-mobility channels is then evaluated.
Simulation results demonstrate a significant performance improvement for OTFS modulation over the conventional orthogonal frequency division multiplexing (OFDM) modulation over high-mobility channels. Analytical results and the effectiveness of the proposed code design are also verified by simulations with the application of both classical and modern codes for OTFS systems.
\end{abstract}

\begin{IEEEkeywords}
OTFS modulation, diversity analysis, code design, high-mobility, fading channels.
\end{IEEEkeywords}

%
\IEEEpeerreviewmaketitle
\section{Introduction}
Beyond the fifth-generation (B5G) wireless communication systems are required to accommodate various emerging applications in high-mobility environments, such as mobile communications on board aircraft (MCA), low-earth-orbit satellites (LEOSs), high speed trains, and unmanned aerial vehicles (UAVs) \cite{Meyer2019road,giambene2018satellite,cai2020joint}.
Currently deployed orthogonal frequency division multiplexing (OFDM) modulation may not be able to support efficient and reliable communications in such scenarios \cite{hwang2008ofdm}.
Therefore, as a potential solution to supporting heterogeneous requirements of B5G wireless systems in high-mobility scenarios, the recently proposed orthogonal time frequency space (OTFS) modulation has attracted substantial attention \cite{Hadani2017orthogonal}.

%
%
%
%

In high-mobility scenarios, wireless channels are usually doubly-dispersive in the time-frequency (TF) domain \cite{hlawatsch2011wireless,tse2005fundamentals}. In specific, the time dispersion is caused by the effect of multi-path, while the frequency dispersion is caused by the Doppler shifts.
Conventionally, OFDM modulation can efficiently mitigate the intersymbol interference (ISI) induced by the time dispersion by introducing a cyclic prefix (CP). However, the success of OFDM modulation relies deeply on maintaining the orthogonality among all the sub-carriers. Note that perfect orthogonality is highly impractical at the receiver, especially in high-mobility environments, due to the exceedingly large frequency dispersion, and
consequently, the performance of conventional OFDM systems is unsatisfactory in such scenarios \cite{hwang2008ofdm}. On the other hand, by invoking the two-dimensional (2D) symplectic finite Fourier
transform (SFFT), OTFS modulates the information symbols in the delay-Doppler (DD) domain, where the channel parameters are relatively stable compared to those in the TF domain \cite{hlawatsch2011wireless}. More importantly, it can be shown that by modulating the information symbols in the DD domain rather than the TF domain, each symbol principally experiences the whole fluctuations of the TF channel over an OTFS frame and thus OTFS modulation offers the potential of exploiting the full channel diversity, achieving a better error performance compared to that of the conventional OFDM modulation in a high-mobility environment~\cite{Hadani2017orthogonal}.
Furthermore, with the domain transformation performed in OTFS modualtion, one can represent the high dynamic channel parameters in the TF domain equivalently by a sparse presentation in the DD domain. This unique property suggests that the acquisition of channel state information (CSI) for OTFS systems can be performed with a low pilot signaling overhead and that the symbol detection for OTFS systems can be carried out with a low complexity \cite{Raviteja2019embedded,Raviteja2018interference}.

To unleash the potential of OTFS modulation, some recent works have focused on the implementation of practical OTFS systems.
For example, a low-complexity modem structure for OTFS systems was proposed in \cite{farhang2017low}. In particular, it showed that the OTFS modulation can be efficiently implemented with simple pre- and post-processing units based on the conventional OFDM modulator.
Besides, to reduce the signaling overhead, an embedded pilot-aided channel estimation method for OTFS modulation was proposed in \cite{Raviteja2019embedded}. By taking advantage of the sparse representation of the wireless channel in the DD domain, this channel estimation method only requires one pilot symbol with a small number of guard symbols in the DD domain.
Furthermore, in order to reduce the detection complexity, various variations of sum-product algorithms (SPAs) have been proposed to facilitate the OTFS detection. In specific, Raviteja \emph{et. al}  proposed an SPA-based OTFS detector, where the inter-Doppler interference (IDI) was approximated as a Gaussian variable to reduce the detection complexity \cite{Raviteja2018interference}.
A variational Bayes detector for OTFS systems was proposed in \cite{Yuan2020simple}. The basic idea of this detector is to approximate the corresponding \emph{a posteriori} distribution of the optimal detection by exploiting the Kullback-Leibler (KL) divergence such that the SPA can be implemented based on a simpler factor graph compared to that of the original OTFS modulation.

Although the aforementioned excellent works have provided guidelines for practical OTFS system designs, the theoretical error performance advantages of OTFS systems over the conventional OFDM systems
have not been thoroughly studied yet, especially for coded cases.
We note that there are several previous works \cite{Raviteja2019effective,Biglieri2019error,Surabhi2019on} on the error performance analysis of OTFS systems. However, these works mainly
considered the uncoded case and their analysis may not be directly extended to the coded cases.
Consequently, the error performance analysis of coded OTFS systems is still missing in the literature to the best of our knowledge.
As commonly recognized, channel coding is an efficient tool to combat fading and channel impairments and thus is a key enabler for reliable communications between users with high-mobility \cite{wu2016survey}. For OTFS modulation, the 2D transformation from the TF domain to the DD domain provides a potential of exploiting the full TF diversity. In this case, a good channel code needs to couple the coded symbols to the 2D OTFS modulation, in order to exploit the full diversity and in the meantime maximize the coding gain.
However, it is still unknown what is the key coding parameter determining the coding gain for OTFS modulation.
Therefore, in order to facilitate the implementation of practical OTFS systems, the error performance analysis for coded OTFS systems needs to be investigated.

In this paper, we aim to analyze the error performance of coded OTFS systems. To this end,
we start from the study of conditional pairwise-error probability (PEP) \cite{Tarokh1998space,vucetic2003space} for a given channel realization.
In order to obtain an accurate performance analysis, we consider two cases of the OTFS transmission depending on the number of independent resolvable paths of the channel and derive the corresponding conditional PEPs.
Since the exact unconditional PEP is generally intractable \cite{Lu2000space},
we resort to the application of some proper bounding techniques to study the conditional PEP and derive the unconditional performance upper bounds.
Based on the unconditional performance bound, the impact of channel coding parameters on the performance of OTFS modulation is unveiled. In particular, we find that the squared Euclidean distance between a pair of codewords is the key parameter that determines the coding gain for coded OTFS systems, given the number of independent resolvable paths.
Therefore, the code design criterion is formulated to optimize the coding gain by maximizing the minimum squared Euclidean distance between all codeword pairs.
The main contributions of this paper can be summarized as follows.
\begin{itemize}
\item We investigate the conditional PEP of OTFS systems for a given channel realization by studying the pairwise Euclidean distance between OTFS codewords. Based on the conditional PEP, we derive the unconditional performance upper bounds for OTFS systems, according to the number of independent resolvable paths. We also show a few properties on the trace and determinant of the codeword difference matrix.
    Furthermore, according to the derived bounds, we define the coding gain and diversity gain of OTFS systems. More importantly, we show that the coding gain of OTFS systems depends on the squared Euclidean distance and the number of independent resolvable paths.
\item According to the derived performance bounds, we show that there is a fundamental trade-off between the diversity gain and the coding gain for OTFS systems. In particular, the diversity gain of OTFS systems improves with the number of independent resolvable paths, while the coding gain declines.
\item Based on the derived performance bounds, we propose our code design criterion to optimize the coding gain, which is to maximize the minimum squared Euclidean distance of among all codeword pairs.
     In order words, traditional good codes with a large minimum Euclidean distance can be directly applied to OTFS systems.
\item We demonstrate a significant performance improvement achieved by the coded OTFS modulation over the coded OFDM modulation over high-mobility channels by numerical simulations. We also provide numerical results of coded OTFS systems over high-mobility channels with various channel codes, such as classical convolutional codes and state-of-art low-density parity-check (LDPC) codes. Our performance analysis and code design are explicitly verified by these results.
\end{itemize}

The rest of this paper is organized as follows. We provide a brief overview and the system model of OTFS modulation in Section II.
In Section III, the derivation of error performance bounds and the code design criterion for coded OTFS systems are investigated.
The numerical results of coded OTFS systems are presented in Section IV and finally a summary is provided in Section V.

\emph{Notations:} The blackboard bold letter ${\mathbb{A}}$ and ${\mathbb{H}}$ denote the constellation set and an arbitrary subspace, respectively; The notations $(\cdot)^{\rm{T}}$, $(\cdot)^{*}$, $\left\| {\cdot} \right\|$, $(\cdot)^{-1}$, and $(\cdot)^{\rm{H}}$ denote the transpose, the conjugate, the Euclidean norm, the inverse, and the Hermitian operations for a matrix, respectively; $\mathbb{E}[\cdot]$ denote the expectation; $\textrm{det}(\cdot)$, $\textrm{tr}(\cdot)$, and $\textrm{vec}(\cdot)$ denote the determinant, the trace, and the vectorization operation; $\textrm{diag}{\{\cdot\}}$ denotes the diagonal matrix; ``$ \otimes $" denotes the Kronecker product operator; ${{{\bf{F}}_N}}$ and ${{{\bf{I}}_M}}$ denote the discrete Fourier transform (DFT) matrix of size $N\times N$ and the identity matrix of size $M\times M$, respectively; $Q(\cdot)$ denotes the tail distribution function of the standard normal distribution, $\delta(\cdot)$ denotes the Dirac delta function, ${I_0}\left( {\cdot} \right)$ denotes the zero-order modified Bessel function of the first kind, respectively; $P(\cdot)$ denotes the probability and $p(\cdot)$ denotes the probability density function (PDF); $f(\cdot)$ denotes an arbitrary function; ${\left( {f(\cdot)} \right)_{\max }}$ and ${\left( {f(\cdot)} \right)_{\min }}$ denote the maximum and minimum values of function $f(\cdot)$, respectively.

\section{OTFS System Model}
In this section, we first review the OTFS concept and then introduce the considered system model.
\subsection{Coded OTFS System Model}
\begin{figure}
\centering
\includegraphics[width=0.7\textwidth]{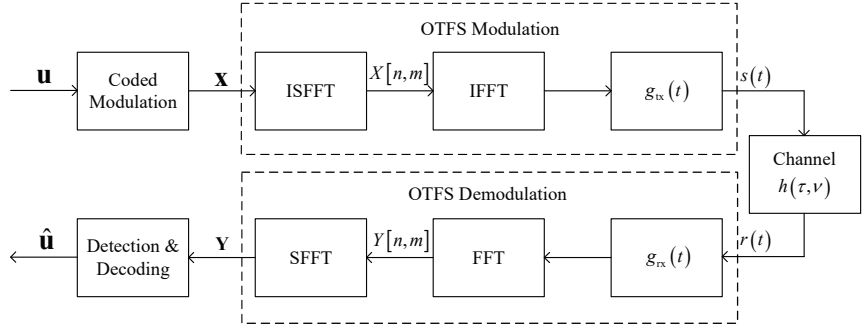}
\caption{The block diagram of an OTFS system.}
\label{System_Model}
\centering
\end{figure}
Without loss of generality, we consider a coded OTFS system as shown in Fig. \ref{System_Model}.
Let $M$ be the number of sub-carriers and $N$ be the number of time slots for each OTFS symbol, respectively.
An information sequence $\bf{u}$ is channel-encoded and then modulated into ${\bf{x}} \in {{\mathbb{A}}^{M N}}$ with length $MN$. Let us arrange ${\bf{x}}$ into a 2D matrix ${\bf{X}} \in {{\mathbb{A}}^{M \times N}}$, i.e., ${\bf{x}} \buildrel \Delta \over = \textrm{vec}\left( {\bf{X}} \right)$, representing the symbols in the DD domain, whose $(k,l)$-th element $x\left[ {k,l} \right]$ is the modulated signal in the $k$-th Doppler and $l$-th delay grid \cite{Hadani2017orthogonal}, for $0 \le k \le N-1,0 \le l \le M-1$.
Each transmitted symbol in the TF domain $X\left[ {n,m} \right], 0 \le n \le N-1, 0 \le m \le M-1$, is then obtained based on ${\bf{X}}$ via the inverse symplectic fast Fourier transform (ISFFT) as follows \cite{Hadani2017orthogonal}
\begin{equation}
X\left[ {n,m} \right] = \frac{1}{{\sqrt {NM} }}\sum\limits_{k = 0}^{N - 1} {\sum\limits_{l = 0}^{M - 1} {x\left[ {k,l} \right]} } {e^{j2\pi \left( {\frac{{nk}}{N} - \frac{{ml}}{M}} \right)}}  .
\end{equation}

\begin{figure}
\centering
\includegraphics[width=0.8\textwidth]{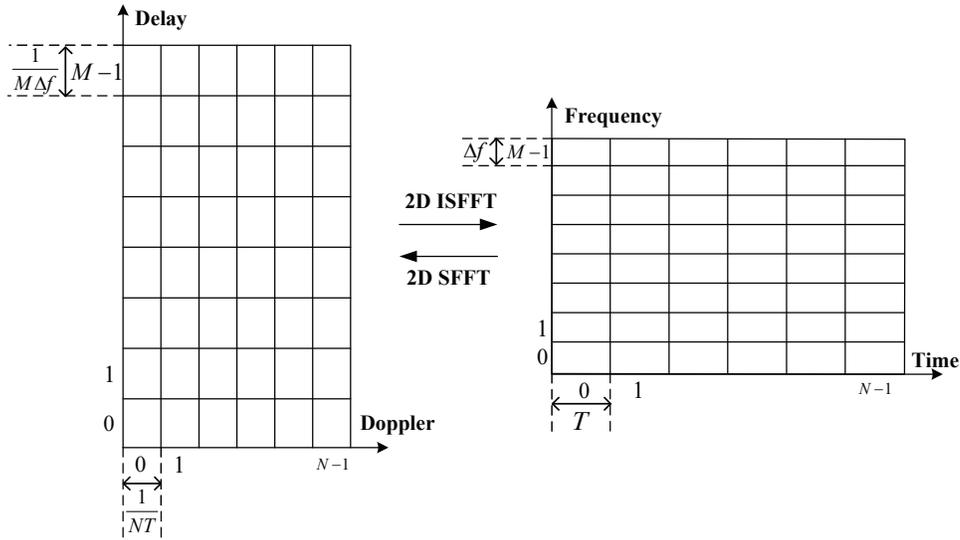}
\caption{The diagram of the transformation between the DD domain and the TF domain.}
\label{Grid}
\centering
\end{figure}

A brief diagram regarding the DD and TF domain transformation is shown in Fig. \ref{Grid}, where $\Delta f$ is the frequency spacing between adjacent sub-carriers and $T = {1 \mathord{\left/
 {\vphantom {1 {\Delta f}}} \right.
 \kern-\nulldelimiterspace} {\Delta f}}$ is the corresponding TF domain time slot duraion.
 Therefore, each OTFS system takes the total bandwidth of $M \Delta f$ and symbol duration $NT$.
In particular, the sampling time $1/(M \Delta  T)$ and sampling frequency $1/(N T)$ are referred to as the \emph{delay resolution} and the \emph{Doppler resolution} of the DD grid, respectively \cite{Raviteja2018interference}, which indicate how precise the acquisition of the channel delay and Doppler can be for the underlying OTFS system.
After the domain transformation (ISFFT), the TF domain transmitted symbol $X\left[ {n,m} \right]$ is then modulated via a conventional OFDM modulator.
The time domain OTFS signal $s\left( t \right)$ is written as
\begin{equation}
s\left( t \right) = \sum\limits_{n = 0}^{N - 1} {\sum\limits_{m = 0}^{M - 1} {X\left[ {n,m} \right]{g_{{\rm{tx}}}}\left( {t - nT} \right){e^{j2\pi m\Delta f\left( {t - nT} \right)}}} },
\end{equation}
where the $g_{{\rm{tx}}}(t)$ denotes the pulse shaping filter.
As described above, for OFDM-based OTFS implementation, the OTFS modulator can be viewed as a concatenation of a precoder (ISFFT) and a conventional OFDM modulator \cite{farhang2017low}, where the OFDM modulator consists of an inverse FFT (IFFT) block and a pulse shaping filter $g_{{\rm{tx}}}(t)$.
Similar to \cite{Hadani2017orthogonal}, we consider the DD domain representation of the time-varying channel, where the channel impulse response is given by
\begin{equation}
h\left( {\tau ,\nu } \right) = \sum\limits_{i = 1}^P {{h_i}\delta \left( {\tau  - {\tau _i}} \right)\delta \left( {\nu  - {\nu _i}} \right)}.
\label{channel}
\end{equation}
In (\ref{channel}), $P$ is the number of independent resolvable paths while $h_i$, $\tau _i$, and $\nu _i$ are the channel coefficients, delay shifts, and Doppler shifts corresponding to the $i$-th path, respectively.
Let $\bar w\left( t \right)$ be the additive white Gaussian noise process with one-sided power spectral density (PSD) $N_0$. The received signal can be written as
\begin{equation}
r\left( t \right) = \int {\int {h\left( {\tau ,\nu } \right)s\left( {t - \tau } \right)} } {e^{j2\pi \nu \left( {t - \tau } \right)}}d\tau d\nu + \bar w\left( t \right).
\label{received_signal}
\end{equation}
Let $g_{{\rm{rx}}}(t)$ be the filter adopted at the receiver side. The received symbols $Y\left[ {n,m} \right]$ in the TF domain are then obtained by
\begin{equation}
Y\left[ {n,m} \right] = \int {r\left( t \right)g_{{\rm{rx}}}^*\left( {t - nT} \right){e^{ - j2\pi m\Delta f\left( {t - nT} \right)}}} dt.
\label{matched_filter}
\end{equation}
Substituting (\ref{received_signal}) into (\ref{matched_filter}) and after similar manipulations as in \cite{Raviteja2018interference}, (\ref{matched_filter}) can be simplified as
\begin{equation}
Y\left[ {n,m} \right] = \sum\limits_{n' = 0}^{N - 1} {\sum\limits_{m' = 0}^{M - 1} {{H_{n,m}}\left[ {n',m'} \right]} } X\left[ {n',m'} \right]+\bar w \left[ {n,m} \right],
\label{receivded_symbols_TF}
\end{equation}
where $\bar w \left[ {n,m} \right]$ is the corresponding TF domain noise sample and ${H_{n,m}}\left[ {n',m'} \right]$ is the TF domain channel response, given by
\begin{align}
{H_{n,m}}\left[ {n',m'} \right]=& \int {\int {h\left( {\tau ,\nu } \right){A_{{g_{{\rm{tx}}}},{g_{{\rm{rx}}}}}}\Big( {\left( {n - n'} \right)T - \tau ,\left( {m - m'} \right)\Delta f - \nu } \Big)} } \notag \\
&{e^{j2\pi \left( {\nu  + m'\Delta f} \right)\left( {\left( {n - n'} \right)T - \tau } \right)}}{e^{j2\pi \nu n'T}}d\tau d\nu.
\label{TF_channel_coef}
\end{align}
In (\ref{TF_channel_coef}), the function ${A_{{g_{{\rm{tx}}}},{g_{{\rm{rx}}}}}}\left( {{\tau _\Delta },{\nu _\Delta }} \right)$ is the so-called \emph{cross-ambiguity} function, which indicates the interference level between the TF domain symbols due to the channel dispersion and is given by \cite{Raviteja2018interference}
\begin{equation}
{A_{{g_{{\rm{tx}}}},{g_{{\rm{rx}}}}}}\left( {{\tau _\Delta },{\nu _\Delta }} \right) \buildrel \Delta \over = \int {{g_{{\rm{tx}}}}\left( t \right)g_{{\rm{rx}}}^*} \left( {t - {\tau _\Delta }} \right){e^{j2\pi {\nu _\Delta }\Delta ft}}dt.
\end{equation}
Hence, the received symbols $y\left[ {k,l} \right]$ in the DD domain are obtained by performing the SFFT on the TF domain received symbols $Y\left[ {n,m} \right]$ which are written as
\begin{equation}
y\left[ {k,l} \right] = \frac{1}{{\sqrt {NM} }}\sum\limits_{n = 0}^{N - 1} {\sum\limits_{m = 0}^{M - 1} {Y\left[ {n,m} \right]{e^{ - j2\pi \left( {\frac{{nk}}{N} - \frac{{ml}}{M}} \right)}}} } + w\left[ {k,l} \right],
\label{DD_y}
\end{equation}
where $w\left[ {k,l} \right]$ denotes the equivalent AWGN samples in the DD domain.
In specific, the DD domain received symbols $y\left[ {k,l} \right]$ can be arranged into the 2D received symbol matrix ${\bf{Y}}$ according to the DD grid,
whose $(k,l)$-th element is $y\left[ {k,l} \right]$.
For the ease of presentation and analysis, we consider the vector form representation of the input-output relationship of OTFS system in the DD domain based on (\ref{DD_y}) in the sequel.

\subsection{Vector Form Representation of OTFS}
Let ${\bf{x}} \buildrel \Delta \over = \textrm{vec}\left( {\bf{X}} \right) \in {{\mathbb{A}}^{MN}}$ and ${\bf{y}} \buildrel \Delta \over = \textrm{vec}\left( {\bf{Y}} \right) \in {{\mathbb{A}}^{MN}}$ denote the vector forms of the transmitted symbols ${\bf{X}}$ and the received symbols ${\bf{Y}}$ in the DD domain, respectively.
According to (\ref{DD_y}), we have
\begin{equation}
{\bf{y}} = {{\bf{H}}_{{\rm{eff}}}}{\bf{x}} + {\bf{w}},\label{system_model_original}
\end{equation}
where $\bf{w}$ is the corresponding noise vector and ${{\bf{H}}_{{\rm{eff}}}}$ of size $MN \times MN$ is the \emph{effective channel matrix} in the DD domain.
Assuming that both $g_{{\rm{tx}}}(t)$ and $g_{{\rm{rx}}}(t)$ are rectangular pulses, with a reduced CP frame format, the effective channel matrix ${{\bf{H}}_{{\rm{eff}}}}$ is given by \cite{Raviteja2019practical}
\begin{equation}
{{\bf{H}}_{{\rm{eff}}}} = \sum\limits_{i = 1}^P {{h_i}}\left( {{{\bf{F}}_N} \otimes {{\bf{I}}_M}} \right) {{\bm{\Pi }}^{{l_i}}}{{\bm{\Delta}} ^{{k_i}}}\left( {{\bf{F}}_N^{\rm{H}} \otimes {{\bf{I}}_M}} \right), \label{channel1}
\end{equation}
where ${\bm{\Pi }}$ is the permutation matrix (forward cyclic shift), i.e.,
\begin{equation}
{\bm{\Pi }} = {\left[ {\begin{array}{*{20}{c}}
0& \cdots &0&1\\
1& \ddots &0&0\\
 \vdots & \ddots & \ddots & \vdots \\
0& \cdots &1&0
\end{array}} \right]_{MN \times MN}},
\end{equation}
and ${\bm{\Delta}}=\textrm{diag}\{z^0,z^1,...,z^{MN-1}\} $ is a diagonal matrix with $z \buildrel \Delta \over = {e^{\frac{{j2\pi }}{{MN}}}}$ \cite{Raviteja2019practical}.
In (\ref{channel1}), $l_i$ and $k_i$ are the delay and Doppler indices corresponding to the $i$-th path, respectively, and we have
\begin{equation}
{\tau _i} = \frac{{l_i}}{{M\Delta f}},\quad
{\nu _i} = \frac{{{k_i} + {\kappa _i}}}{{NT}}.
\label{resolution}
\end{equation}
Note that the term $- {1 \mathord{\left/
 {\vphantom {1 2}} \right.
 \kern-\nulldelimiterspace} 2} \le {\kappa _i} \le {1 \mathord{\left/
 {\vphantom {1 2}} \right.
 \kern-\nulldelimiterspace} 2}$ denotes the fractional Doppler which corresponds to the fractional shift from the nearest Doppler \cite{Raviteja2018interference}. On the other hand, since the typical value of the sampling time ${1 \mathord{\left/
 {\vphantom {1 {M\Delta f}}} \right.
 \kern-\nulldelimiterspace} {M\Delta f}}$ is in the delay domain usually sufficiently small, the impact of
fractional delays in typical wide-band systems can be neglected \cite{tse2005fundamentals}.
It should be noted that (\ref{channel1}) still holds even for the fractional Doppler case, i.e., ${\kappa _i} \ne 0$, due to the properties of diagonal matrix $\bm{\Delta}$.
For simplicity, we only consider the integer Doppler case in this paper.
To facilitate the following analysis, we assume that the delay and Doppler indices $l_i$ and $k_i$ follow the discrete uniform distribution.
In specific, we have $l_i \in \left[ {0,{l_{\max }}} \right]$, where $l_{\max }$ is the maximum delay index.
We also have $k_i \in \left[ { - {k_{\max }},{k_{\max }}} \right]$, where $k_{\max }$ is the maximum Doppler index\footnote{The maximum Doppler shift is given by ${\nu_{\max }} = \frac{v}{c}{f_c}$, where $v$ is the relative user equipment (UE) speed, $c$ is the speed of light, and $f_c$ is the carrier frequency, respectively. Thus, the maximum Doppler index is given by ${k_{\max }} = {\nu_{\max }} N T$ \cite{Raviteja2018interference}. The maximum delay shift is given by ${\tau_{\max }} = \frac{d_{\rm{max}}}{c}$, where $d_{\rm{max}}$ is the maximum distance among the $P$ channel paths.}.
In particular, let ${{\bm{\omega}} _\tau }$ and ${{\bm{\omega}} _\nu }$ denote the vectors of delay indices and Doppler indices, respectively, i.e., ${{\bm{\omega}} _\tau } = \left[ {{l_1},{l_2}, \ldots ,{l_P}} \right]^{\rm{T}}$, ${{\bm{\omega}} _\nu } = \left[ {{k_1},{k_2}, \ldots ,{k_P}} \right]^{\rm{T}}$, respectively.
Therefore, according to \cite{Raviteja2019effective}, (\ref{system_model_original}) can be rewritten as
\begin{align}
{\bf{y}} = {\bf{\Phi }}_{{{\bm{\omega}} _\tau },{{\bm{\omega}} _\nu }}\left( {\bf{x}} \right){\bf{h}} + {\bf{w}} \label{system_model},
\end{align}
where ${\bf{\Phi }}_{{{\bm{\omega}} _\tau },{{\bm{\omega}} _\nu }}\left( {\bf{x}} \right)$ is referred to as the \emph{equivalent codeword matrix} and it is a concatenated matrix of size $MN\times P$ constructed by the column vector ${{\bf{\Xi }}_i}{\bf{x}}$ , i.e.,
\begin{equation}
{\bf{\Phi }}_{{{\bm{\omega}} _\tau },{{\bm{\omega}} _\nu }}\left( {\bf{x}} \right) = \left[ {{{\bf{\Xi }}_1}{\bf{x}}\quad {{\bf{\Xi }}_2}{\bf{x}}\quad  \cdots \quad {{\bf{\Xi }}_P}{\bf{x}}} \right],
\label{equation_phi}
\end{equation}
and ${{\bf{\Xi }}_i}$ is given by
\begin{equation}
{{\bf{\Xi }}_i} \buildrel \Delta \over = \left( {{{\bf{F}}_N} \otimes {{\bf{I}}_M}} \right){{\bm{\Pi }}^{{l_i}}}{{\bm{\Delta}} ^{{k_i}}}\left( {{\bf{F}}_N^{\rm{H}} \otimes {{\bf{I}}_M}} \right), 1 \le i \le P.
\label{equation_Xi}
\end{equation}
In (\ref{system_model}), $\bf{h}$ is the path coefficient vector of size $P\times 1$, i.e., ${\bf{h}} = {\left[ {{h_1},{h_2},...,{h_P}} \right]^{\rm{T}}}$, where the elements in ${\bf{h}}$
are assumed to be independent and identically distributed complex Gaussian random variables.
Besides, we assume that $h_i$ has the mean $\mu $ and the variance $1/(2P)$ per real dimension\footnote{Here, we assume a uniform power delay profile of the channel.}, for $1 \le i \le P$.
In particular, we note that if $\mu=0$, $\left| {{h_i}} \right|$ follows the Rayleigh distribution, which will be considered as a special case in our error performance analysis and code design.
Based on (\ref{system_model}), the error performance analysis of the OTFS systems will be conducted in the next section.

\section{Error Performance Analysis}
In order to investigate the error performance of the coded OTFS systems, we assume that ideal channel state information (CSI) is available at the receiver, including $\bf{h}$, ${{\bm{\omega}} _\tau }$, and ${{\bm{\omega}} _\nu }$ \cite{Raviteja2019embedded,Yuan2020Orthogonal_report}.
We note that matrix ${\bf{\Phi }}_{{{\bm{\omega}} _\tau },{{\bm{\omega}} _\nu }}\left( {\bf{x}} \right)$ depends on ${{\bm{\omega}} _\tau }$, ${{\bm{\omega}} _\nu }$, and the transmitted symbol vector ${\bf{x}}$. Therefore, for a given channel realization, we define the \emph{conditional Euclidean distance} ${d_{\bf{h},{{\bm{\omega}} _\tau },{{\bm{\omega}} _\nu }}^2\left( {{\bf{x}},{\bf{x'}}} \right)}$ between a pair of codewords ${\bf{x}}$ and ${\bf{x'}}$ (${\bf{x}} \ne {\bf{x'}}$) as
\begin{equation}
d_{\bf{h},{{\bm{\omega}} _\tau },{{\bm{\omega}} _\nu }}^2\left( {{\bf{x}},{\bf{x'}}} \right) = d_{\bf{h},{{\bm{\omega}} _\tau },{{\bm{\omega}} _\nu }}^2\left( {\bf{e}} \right)\buildrel \Delta \over ={\left\| {{{\bf{\Phi }}_{{{\bm{\omega}} _\tau },{{\bm{\omega}} _\nu }}\left( {\bf{e}} \right)}{\bf{h}}} \right\|^2}={{\bf{h}}^{\rm{H}}}{\bf{\Omega }}_{{{\bm{\omega}} _\tau },{{\bm{\omega}} _\nu }}\left( {\bf{e}} \right){\bf{h}},
\end{equation}
where ${\bf{e}} = {\bf{x}} - {\bf{x'}}$ is the corresponding codeword difference (error) sequence and ${\bf{\Omega }}_{{{\bm{\omega}} _\tau },{{\bm{\omega}} _\nu }}\left( {\bf{e}} \right) = {\left( {{\bf{\Phi }}_{{{\bm{\omega}} _\tau },{{\bm{\omega}} _\nu }}\left( {\bf{e}} \right)} \right)^{\rm{H}}}\left( {\bf{\Phi }}_{{{\bm{\omega}} _\tau },{{\bm{\omega}} _\nu }}\left( {\bf{e}} \right)\right)$ is referred to as the \emph{codeword difference matrix}. Without loss of generality and for notational simplicity,
we henceforth drop the subscript of ${\bf{\Omega }}_{{{\bm{\omega}} _\tau },{{\bm{\omega}} _\nu }}\left( {\bf{e}} \right)$ and we now have
\begin{equation}
{\bf{\Omega }}\left( {\bf{e}} \right) = \left[ {\begin{array}{*{20}{c}}
{{{\bf{e}}^{\rm{H}}}{\bf{\Xi }}_1^{\rm{H}}{{\bf{\Xi }}_1}{\bf{e}}}&{{{\bf{e}}^{\rm{H}}}{\bf{\Xi }}_1^{\rm{H}}{{\bf{\Xi }}_2}{\bf{e}}}& \cdots &{{{\bf{e}}^{\rm{H}}}{\bf{\Xi }}_1^{\rm{H}}{{\bf{\Xi }}_P}{\bf{e}}}\\
{{{\bf{e}}^{\rm{H}}}{\bf{\Xi }}_2^{\rm{H}}{{\bf{\Xi }}_1}{\bf{e}}}&{{{\bf{e}}^{\rm{H}}}{\bf{\Xi }}_2^{\rm{H}}{{\bf{\Xi }}_2}{\bf{e}}}&{}& \vdots \\
 \vdots &{}& \ddots & \vdots \\
{{{\bf{e}}^{\rm{H}}}{\bf{\Xi }}_P^{\rm{H}}{{\bf{\Xi }}_1}{\bf{e}}}& \cdots & \cdots &{{{\bf{e}}^{\rm{H}}}{\bf{\Xi }}_P^{\rm{H}}{{\bf{\Xi }}_P}{\bf{e}}}
\end{array}} \right]. \label{Gram}
\end{equation}
The conditional PEP is upper-bounded by \cite{Tarokh1998space,vucetic2003space}
\begin{equation}
P\left( {\left. {{\bf{x}},{\bf{x'}}} \right|{\bf{h}},{{{\bm{\omega}} _\tau }},{{{\bm{\omega}} _\nu }}} \right) \le \exp \left( { - \frac{{{E_s}}}{{4{N_0}}}{d_{\bf{h},{{\bm{\omega}} _\tau },{{\bm{\omega}} _\nu }}^2\left( {{\bf{x}},{\bf{x'}}} \right)}} \right), \label{PEP_derivation1}
\end{equation}
where $E_s$ is the average symbol energy.
Note that the codeword difference matrix ${\bf{\Omega }}\left( {\bf{e}} \right)$ is nonnegative definite Hermitian.
Let us denote by $\left\{ {{{\bf{v}}_1},{{\bf{v}}_2},...,{{\bf{v}}_P}} \right\}$ the eigenvectors of ${\bf{\Omega }}\left( {\bf{e}} \right)$ and
$\left\{ {{\lambda _1},{\lambda _2},...,{\lambda _P}} \right\}$ the corresponding nonnegative real eigenvalues sorted in the descending order.
Thus, (\ref{PEP_derivation1}) can be further expanded as \cite{Tarokh1998space,vucetic2003space}
\begin{equation}
P\left( {{\left. {{\bf{x}},{\bf{x'}}} \right|{\bf{h}},{{{\bm{\omega}} _\tau }},{{{\bm{\omega}} _\nu }}}} \right) \le \exp \left( { - \frac{{{E_s}}}{{4{N_0}}}\sum\limits_{i = 1}^r {{\lambda _i}{{\left| {{{\tilde h}_i}} \right|}^2}} } \right), \label{PEP_derivation2}
\end{equation}
where $r$ is the rank of ${\bf{\Omega }}\left( {\bf{e}} \right)$, i.e., $r \le P$, and ${{\tilde h}_i} = {\bf{h}}\cdot{{\bf{v}}_i}$, for $1 \le i \le r$.
It can be shown that $\left\{ {{{\tilde h}_1},{{\tilde h}_2},...,{{\tilde h}_r}} \right\}$ are independent complex Gaussian random variables with mean ${\mu _{{{\tilde h}_i}}}=\mathbb{E}\left[ {\bf{h}} \right]\cdot{{\bf{v}}_i}$ and variance $1/(2P)$ per real dimension.
Thus, it is obvious that ${\small| {{{\tilde h}_i}} \small|}$ follows the Rician distribution with Rician factor ${K_i} = {\left| {{\mu _{{{\tilde h}_i}}}} \right|^2}$ \cite{Tarokh1998space}, and its
PDF is given by
\begin{equation}
p\left( {\left| {{{\tilde h}_i}} \right|} \right) = 2P\left| {{{\tilde h}_i}} \right|\exp \left( { - P{{\left| {{{\tilde h}_i}} \right|}^2} - P{K_i}} \right){I_0}\left( {2P\left| {{{\tilde h}_i}} \right|\sqrt {{K_i}} } \right).\label{Rician_PDF}
\end{equation}
In the following, we will target on the analysis of the unconditional PEP. To this end, we aim to calculate the average of (\ref{PEP_derivation2}) over the channel distribution according to (\ref{Rician_PDF}), and consider the impact of the distributions of delay and Doppler indices. More specifically, we will discuss two important cases depending on the number of independent resolvable paths $P$.

\textbf{Remarks}: It has been defined in the previous works \cite{Raviteja2019effective,Biglieri2019error,Surabhi2019on} that the rank of ${\bf{\Omega }}\left( {\bf{e}} \right)$ is the \textbf{diversity} gain of the OTFS modulation. Specifically, it has been shown in \cite{Surabhi2019on} that the diversity gain of uncoded OTFS modulation systems can be one but the full diversity can be obtained by suitable precoding schemes. Furthermore, \cite{Raviteja2019effective} has shown that the full diversity can be achieved almost surely when the frame size is sufficiently large, even for uncoded OTFS modulation systems.
Therefore, in the following, we will mainly focus on the analysis of the coding gain when the OTFS modulation achieves the full diversity, i.e., $r=P$.

\subsection{Error Performance Analysis for Coded OTFS systems}
Notice that $\bf{h}$, ${{\bm{\omega}} _\tau }$, and ${{{\bm{\omega}} _\nu }}$ are independent from each other. Therefore, the unconditional PEP can be derived by firstly averaging (\ref{PEP_derivation2}) over ${| {{{\tilde h}_i}} |}$ term by term which results in
\begin{equation}
P\left( {\left. {{\bf{x}},{\bf{x'}}} \right|{{{\bm{\omega}} _\tau }},{{{\bm{\omega}} _\nu }}} \right) \le \prod\limits_{i = 1}^P {\frac{1}{{1 + \frac{{{E_s}}}{{4{N_0}}}\cdot\frac{{{\lambda _i}}}{{ P }}}}\exp \left( { - \frac{{{K_i}\frac{{{E_s}}}{{4{N_0}}}\cdot\frac{{{\lambda _i}}}{{ P }}}}{{1 + \frac{{{E_s}}}{{4{N_0}}}\cdot\frac{{{\lambda _i}}}{{ P }}}}} \right)}  . \label{PEP_derivation3}
\end{equation}
Furthermore, we consider a special case where ${{K_i}}=0$ and ${\small| {{{\tilde h}_i}} \small|}$ follows the Rayleigh distribution, i.e., ${\small| {{{h}_i}} \small|}$ also follows the Rayleigh distribution.
In the case of \textbf{Rayleigh fading}, (\ref{PEP_derivation3}) can be further simplified as
\begin{equation}
P\left( {\left. {{\bf{x}},{\bf{x'}}} \right|{{{\bm{\omega}} _\tau }},{{{\bm{\omega}} _\nu }}} \right)\le {\left( {\prod\limits_{i = 1}^P {{\lambda _i}/P} } \right)^{ - 1}}{\left( {\frac{{{E_s}}}{{4{N_0}}}} \right)^{ - P}}. \label{PEP_derivation4}
\end{equation}
By noticing that the term $\prod\nolimits_{i = 1}^P {{\lambda _i}} $ equals to the determinant of ${\bf{\Omega }}\left( {\bf{e}} \right)$,
(\ref{PEP_derivation4}) can be written as
\begin{equation}
P\left( {\left. {{\bf{x}},{\bf{x'}}} \right|{{{\bm{\omega}} _\tau }},{{{\bm{\omega}} _\nu }}} \right)\le\frac{1}{{\det \left( {{\bf{\Omega }}\left( {\bf{e}} \right)} \right)}}{\left( {\frac{{{E_s}}}{{4{N_0}P}}} \right)^{ - P}}. \label{New_PEP_derivation1}
\end{equation}
It should be noted that (\ref{New_PEP_derivation1}) is consistent with the analysis in \cite{Biglieri2019error}. On the other hand, the PEP in (\ref{New_PEP_derivation1}) depends on the delay and Doppler indices ${\bm{\omega}} _\tau$ and ${\bm{\omega}} _\nu$.
In order to derive the unconditional PEP, we need to find the statistical distribution for the determinant of ${\bf{\Omega }}\left( {\bf{e}} \right)$ regarding the delay and Doppler indices.
Unfortunately, such a task is generally intractable \cite{Lu2000space} and is normally handled by applying the Monte Carlo method without providing any important insight.
Instead of resorting to the Monte Carlo method, we apply proper bounding techniques to evaluate the determinant of ${\bf{\Omega }}\left( {\bf{e}} \right)$ in order to obtain some general results about the unconditional PEP.
In particular, we assume that ${\bf{\Omega }}\left( {\bf{e}} \right)$ is of full-rank for any channel parameters ${{\bm{\omega}} _\tau }$ and ${{{\bm{\omega}} _\nu }}$ and error sequence \cite{Raviteja2019effective,Surabhi2019on}, in which case we have the following property.

\textbf{Property 1} \emph{(Gram matrix \cite{Tut_Gram})}:
Let ${\bar{\bf{u}}_i} \buildrel \Delta \over = {{\bf{\Xi }}_i}{\bf{e}}$, for $1\le i\le P$, where ${\bf{\Xi }}_i$ is given by (\ref{equation_Xi}). Then,
$\left\{ {{\bar{\bf{u}}_1},{\bar{\bf{u}}_2}, \ldots {\bar{\bf{u}}_{P}}} \right\}$ form a list of independent vectors chosen from the $P$-dimensional complex inner-product subspace $\mathbb{H}^{P}$.
Thus, the codeword difference matrix ${\bf{\Omega }}\left( {\bf{e}} \right)$ is positive definite Hermitian and it is a \emph{Gram matrix} corresponding to the vectors $\left\{ {{\bar{\bf{{u}}}_1},{\bar{\bf{{u}}}_2}, \ldots {\bar{\bf{{u}}}_P}} \right\}$.

Based on the property of ${\bf{\Omega }}\left( {\bf{e}} \right)$, we now introduce four important Lemmas, which will be served as the building blocks for our error performance analysis for coded OTFS systems.

\textbf{Lemma 1} \emph{(Main diagonal elements of ${\bf{\Omega }}\left( {\bf{e}} \right)$)}:
The main diagonal elements of the codeword difference matrix ${\bf{\Omega }}\left( {\bf{e}} \right)$ are of the same value ${d_{\rm{E}}^2\left( {\bf{e}} \right)}$,
where $d_{\rm{E}}^2\left( {\bf{e}} \right) = {{\bf{e}}^{\rm{H}}}{\bf{e}}$ is the squared Euclidean distance for a pair of codewords ${\bf{x}}$ and ${\bf{x'}}$ with the corresponding to the error sequence $\bf{e}$.

\emph{Proof}: By considering (\ref{equation_phi}), the $i$-th diagonal element of ${\bf{\Omega }}\left( {\bf{e}} \right)$ is given by ${{\bf{e}}^{\rm{H}}}{\bf{\Xi }}_i^{\rm{H}}{{\bf{\Xi }}_i}{\bf{e}}$, and it is equal to the inner product of ${{{{\bf{\bar u}}}_i}}$.
Furthermore, we have
\begin{align}
{\bf{\Xi }}_i^{\rm{H}}{{\bf{\Xi }}_i} =& \left( {{{\bf{F}}_N} \otimes {{\bf{I}}_M}} \right){\left( {{{\bm{\Delta}} ^{{k_i}}}} \right)^{\rm{H}}}{\left( {{{\bm{\Pi }}^{{l_i}}}} \right)^{\rm{H}}}{{\bm{\Pi }}^{{l_i}}}{{\bm{\Delta}} ^{{k_i}}}\left( {{\bf{F}}_N^{\rm{H}} \otimes {{\bf{I}}_M}} \right) \notag\\
=&\left( {{{\bf{F}}_N} \otimes {{\bf{I}}_M}} \right){{\bm{\Delta}} ^{ - {k_i}}}{{\bm{\Pi }}^{ - {l_i}}}{{\bm{\Pi }}^{{l_i}}}{{\bm{\Delta}} ^{{k_i}}}\left( {{\bf{F}}_N^{\rm{H}} \otimes {{\bf{I}}_M}} \right)
\label{omega_derivation1}\\
=&{{\bf{I}}_{MN}} \label{omega_derivation2},
\end{align}
where (\ref{omega_derivation1}) is due to the diagonal properties of ${\bm{\Pi }}$ and $\bm{\Delta}$, respectively, and (\ref{omega_derivation2}) is due to the property of the Kronecker product.
Therefore, the term ${{\bf{e}}^{\rm{H}}}{\bf{\Xi }}_i^{\rm{H}}{{\bf{\Xi }}_i}{\bf{e}}$ is further simplified as ${d_{\rm{E}}^2\left( {\bf{e}} \right)}$.
This completes the proof of Lemma 1.

\textbf{Lemma 2} \emph{(Trace of ${\bf{\Omega }}\left( {\bf{e}} \right)$)}:
The trace of the codeword difference matrix ${\bf{\Omega }}\left( {\bf{e}} \right)$ is ${Pd_{\rm{E}}^2\left( {\bf{e}} \right)}$.

\emph{Proof}: This lemma is a straightforward extension of Lemma 1.

\textbf{Lemma 3} \emph{(Lower bound on the trace of ${{{\left( {{\bf{\Omega }}\left( {\bf{e}} \right)} \right)}^{ - 1}}}$)}:
The trace of the inverse of the codeword difference matrix ${\bf{\Omega }}\left( {\bf{e}} \right)$ is lower-bounded by
\begin{equation}
{{\rm{tr}}\left( {{{\left( {{\bf{\Omega }}\left( {\bf{e}} \right)} \right)}^{ - 1}}} \right)} \ge \frac{P}{{d_{\rm{E}}^2\left( {\bf{e}} \right)}}, \label{Inverse_trace}
\end{equation}
where the equality holds if ${\bf{\Omega }}\left( {\bf{e}} \right)$ is a diagonal matrix, i.e., ${\bf{\Omega }}\left( {\bf{e}} \right)=\textrm{diag}\left\{ {d_{\rm{E}}^2{\left( {\bf{e}} \right)}, \ldots ,d_{\rm{E}}^2}{\left( {\bf{e}} \right)} \right\}$.

\emph{Proof}: The proof is given in Appendix A.

\textbf{Lemma 4} \emph{(Lower bound on the summation of eigenvalue squares of ${{{{{\bf{\Omega }}\left( {\bf{e}} \right)}}}}$)}:
The summation of the squared eigenvalues of ${{{ {{\bf{\Omega }}\left( {\bf{e}} \right)} }}}$ is lower-bounded by
\begin{equation}
\sum\limits_{i = 1}^P {\lambda _i^2}  \ge P{\left( {d_{\rm{E}}^2\left( {\bf{e}} \right)} \right)^2}, \label{Lemma4}
\end{equation}
where the equality holds if ${\bf{\Omega }}\left( {\bf{e}} \right)$ is a diagonal matrix, i.e., ${\bf{\Omega }}\left( {\bf{e}} \right)=\textrm{diag}\left\{ {d_{\rm{E}}^2{\left( {\bf{e}} \right)}, \ldots ,d_{\rm{E}}^2}{\left( {\bf{e}} \right)} \right\}$.

\emph{Proof}: It is obvious that the eigenvalues $\lambda_i$ of the codeword difference matrix ${{\bf{\Omega }}\left( {\bf{e}} \right)}$ are all positive when ${{\bf{\Omega }}\left( {\bf{e}} \right)}$ is full-rank.
Therefore, we apply the Cauchy-Schwarz inequality and the following holds
\begin{equation}
\sum\limits_{i = 1}^P {\lambda _i^2}  \ge \frac{1}{P}{\left( {\sum\limits_{i = 1}^P {{\lambda _i}} } \right)^2} = \frac{1}{P}{\left( {{\rm{tr}}\left( {{\bf{\Omega }}\left( {\bf{e}} \right)} \right)} \right)^2} = P{\left( {d_{\rm{E}}^2\left( {\bf{e}} \right)} \right)^2},
\end{equation}
where the equality is only achieved when the eigenvalues of ${\bf{\Omega }}\left( {\bf{e}} \right)$ are the same, e.g., ${\bf{\Omega }}\left( {\bf{e}} \right)=\textrm{diag}\left\{ {d_{\rm{E}}^2{\left( {\bf{e}} \right)}, \ldots ,d_{\rm{E}}^2}{\left( {\bf{e}} \right)} \right\}$. This completes the proof of Lemma 4.

The above Lemmas show some important properties of the codeword difference matrix ${\bf{\Omega }}\left( {\bf{e}} \right)$. Based on these properties of ${\bf{\Omega }}\left( {\bf{e}} \right)$, we can now consider the following lower bounds of the determinant for the codeword difference matrix ${\bf{\Omega }}\left( {\bf{e}} \right)$.

%

\textbf{Theorem 1} \emph{(Lower bound on the determinant of ${\bf{\Omega }}\left( {\bf{e}} \right)$)}:
The determinant of the codeword difference matrix ${\bf{\Omega }}\left( {\bf{e}} \right)$ is lower-bounded by
\begin{equation}
\det \left( {{\bf{\Omega }}\left( {\bf{e}} \right)} \right) \ge {\left( {d_{\rm{E}}^2} \left( {\bf{e}} \right)\right)^P}\exp \left( {P - d_{\rm{E}}^2\left( {\bf{e}} \right){\rm{tr}}\left( {{{\left( {{\bf{\Omega }}\left( {\bf{e}} \right)} \right)}^{ - 1}}} \right)} \right), \label{determinant_lower_bound}
\end{equation}
where the equality holds if ${\bf{\Omega }}\left( {\bf{e}} \right)$ is a diagonal matrix, i.e., ${\bf{\Omega }}\left( {\bf{e}} \right)=\textrm{diag}\left\{ {d_{\rm{E}}^2{\left( {\bf{e}} \right)}, \ldots ,d_{\rm{E}}^2}{\left( {\bf{e}} \right)} \right\}$.

\emph{Proof}: The proof is given in Appendix B.

It should be noted that (\ref{determinant_lower_bound}) still depends on the channel parameters ${\bm{\omega}} _\tau$ and ${\bm{\omega}} _\nu$.
To obtain an unconditional lower bound, we apply an approximation to the determinant bound which is summarized in the following Theorem.

\textbf{Theorem 2} \emph{(Approximated lower bound on the determinant of ${\bf{\Omega }}\left( {\bf{e}} \right)$)}:
The determinant of the codeword difference matrix ${\bf{\Omega }}\left( {\bf{e}} \right)$ can be approximately lower-bounded by
\begin{equation}
\det \left( {{\bf{\Omega }}\left( {\bf{e}} \right)} \right)  \mathbin{\lower.3ex\hbox{$\buildrel>\over
{\smash{\scriptstyle\sim}\vphantom{_x}}$}} {\left( {d_{\rm{E}}^2\left( {\bf{e}} \right)} \right)^P}, \label{app_determinant_lower_bound}
\end{equation}
where the approximation is exact if ${\bf{\Omega }}\left( {\bf{e}} \right)$ is a diagonal matrix, i.e., ${\bf{\Omega }}\left( {\bf{e}} \right)=\textrm{diag}\left\{ {d_{\rm{E}}^2{\left( {\bf{e}} \right)}, \ldots ,d_{\rm{E}}^2}{\left( {\bf{e}} \right)} \right\}$.

\emph{Proof}: The proof is given in Appendix C.

Note that the approximated lower bound of the determinant of ${\bf{\Omega }}\left( {\bf{e}} \right)$ does not depend on the delay and Doppler indices.
Furthermore, based on Theorem 2, it is not hard to notice that the determinant of the codeword difference matrix ${\bf{\Omega }}\left( {\bf{e}} \right)$ can be approximated by the term ${\left( {d_{\rm{E}}^2\left( {\bf{e}} \right)} \right)^P}$.
In particular, the approximation becomes exact if ${\bf{\Omega }}\left( {\bf{e}} \right)=\textrm{diag}\left\{ {d_{\rm{E}}^2{\left( {\bf{e}} \right)}, \ldots ,d_{\rm{E}}^2}{\left( {\bf{e}} \right)} \right\}$, which can be interpreted as the projections of the error sequence $\bf{e}$ onto each independent resolvable path, i.e., ${\bar{\bf{u}}_i}$, are orthogonal to each other.
According to Theorem 2, we obtain the unconditional PEP as
\begin{equation}
P\left( { {{\bf{x}},{\bf{x'}}} } \right)\mathbin{\lower.3ex\hbox{$\buildrel<\over
{\smash{\scriptstyle\sim}\vphantom{_x}}$}}{\left( {\frac{{d_{\rm{E}}^2\left( {\bf{e}} \right)}}{P}} \right)^{ - P}}{\left( {\frac{{{E_s}}}{{4{N_0}}}} \right)^{ - P}}. \label{New_Unconditional_PEP2}
\end{equation}
Based on (\ref{New_Unconditional_PEP2}), we notice that the unconditional PEP for OTFS modulation depends on ${d_{\rm{E}}^2\left( {\bf{e}} \right)}$ and number of independent resolvable paths $P$.
In particular, the term $P$ in the denominator can be interpreted as the energy averaging with respect to the number of independent paths, while the term ${\left( {d_{\rm{E}}^2\left( {\bf{e}} \right)} \right)^{ - P}}$
indicates the potential improvement of the error performance introduced by channel coding. Furthermore,
according to (\ref{New_Unconditional_PEP2}), we refer to the power of the signal-to-noise ratio (SNR) as the \textbf{diversity gain}, which dominates the exponential behaviour of the error performance for OTFS systems against the average SNR. On the other hand, the term ${{d_{\rm{E}}^2\left( {\bf{e}} \right)} \mathord{\left/
 {\vphantom {{d_{\rm{E}}^2\left( {\bf{e}} \right)} P}} \right.
 \kern-\nulldelimiterspace} P}$ is referred to as the \textbf{coding gain}, which characterizes the approximate improvement of coded OTFS systems over the uncoded counterpart with the same diversity gain, i.e., the same exponent $-P$ \cite{Tarokh1998space}.
It is interesting to see from (\ref{New_Unconditional_PEP2}) that, there exists a fundamental trade-off between the diversity gain and the coding gain which is formally stated in the following.

\textbf{Corollary 1} \emph{(Trade-off between diversity and coding gain)}:
For a given channel code, the diversity gain of OTFS systems improves with the number of independent resolvable paths $P$, while the coding gain declines.


Based on Corollary 1, we note that when $P$ is small, the diversity gain is small. In this case, the squared Euclidean distance between codewords
 is crucial for OTFS systems as an optimized code can greatly improve the error performance.
On the other hand, when $P$ is large, there is a large diversity gain. In this case, it is expected that the code design can only offer a limited error performance improvement. However, it should also be noted that the coding gain always improves with the increase of $d_{\rm{E}}^2\left( {\bf{e}} \right)$, regardless of the value of channel parameter $P$ according to (\ref{New_Unconditional_PEP2}). Therefore, a preliminary guideline for the code design for the OTFS systems is to maximize the minimum value of $d_{\rm{E}}^2\left( {\bf{e}} \right)$ among all pairs of codewords of the code.

To verify the accuracy of the derived unconditional PEP bound, we numerically compare the coding gain corresponding to (\ref{New_Unconditional_PEP2}) and (\ref{New_PEP_derivation1}).
In particular, recalling (\ref{New_PEP_derivation1}), after some manipulations, we obtain
\begin{equation}
P\left( {\left. {{\bf{x}},{\bf{x'}}} \right|{{{\bm{\omega}} _\tau }},{{{\bm{\omega}} _\nu }}} \right)\le{\left( {\frac{{{{\left( {\det \left( {{\bf{\Omega }}\left( {\bf{e}} \right)} \right)} \right)}^{\frac{{\rm{1}}}{P}}}}}{P}} \right)^{ - P}}{\left( {\frac{{{E_s}}}{{4{N_0}}}} \right)^{ - P}}.
\end{equation}
Hence, we refer to the term ${{{{\left( {\det \left( {{\bf{\Omega }}\left( {\bf{e}} \right)} \right)} \right)}^{\frac{{\rm{1}}}{P}}}} \mathord{\left/
 {\vphantom {{{{\left( {\det \left( {{\bf{\Omega }}\left( {\bf{e}} \right)} \right)} \right)}^{\frac{{\rm{1}}}{P}}}} P}} \right.
 \kern-\nulldelimiterspace} P}$ as the \emph{conditional coding gain} of the OTFS systems for given channel parameters ${\bm{\omega}} _\tau$ and ${\bm{\omega}} _\nu$, as ${ {\det \left( {{\bf{\Omega }}\left( {\bf{e}} \right)} \right)} }$ is a function of ${{{\bm{\omega}} _\tau }},{{{\bm{\omega}} _\nu }}$.
We can also obtain the average coding gain with respect to various channel parameters ${{{\bm{\omega}} _\tau }},{{{\bm{\omega}} _\nu }}$ and error sequences $\bf{e}$.
On the other hand, from the unconditional PEP upper bound (\ref{New_Unconditional_PEP2}), we call the function $f\left( {d_{\rm{E}}^2\left( {\bf{e}} \right)} \right) = {{d_{\rm{E}}^2\left( {\bf{e}} \right)} \mathord{\left/
 {\vphantom {{d_{\rm{E}}^2\left( {\bf{e}} \right)} P}} \right.
 \kern-\nulldelimiterspace} P}$ the \emph{coding gain bound} of the OTFS systems.
Now let us compare the average coding gain and the coding gain obtained from the performance bound via simulations.
In particular, we consider a binary phase shift keying (BPSK) signal for the OTFS system with $M=2$ and $N=5$, and error sequences with ${d_{\rm{E}}^2\left( {\bf{e}} \right)}$, where the maximum delay $l_{\rm{max}}$ and Doppler index $k_{\rm{max}}$ are set to be $2$ and $4$, respectively. To be more specific, we numerically average the
conditional coding gains subjected to all error sequences with ${d_{\rm{E}}^2\left( {\bf{e}} \right)}$ and channel parameters ${{{\bm{\omega}} _\tau }},{{{\bm{\omega}} _\nu }}$ to obtain the average coding gain.
\begin{figure}
\centering
\includegraphics[width=0.6\textwidth]{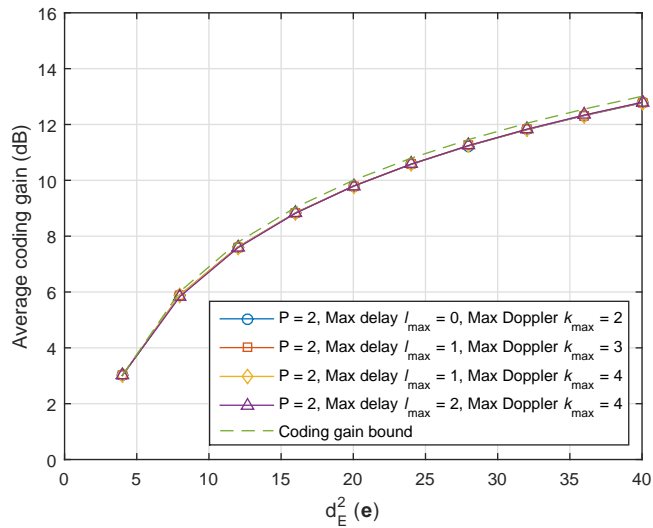}
\caption{Average coding gain for error sequences with $P=2$ and ${d_{\rm{E}}^2\left( {\bf{e}} \right)}$ in terms of various delay and Doppler indices, comparing with the coding gain bound.}
\label{Average_coding_gain_curve1}
\centering
\end{figure}
Fig. \ref{Average_coding_gain_curve1} shows the comparison between the average coding gain and the corresponding coding gain bound in decibels (dB) for $P=2$. As shown in the figure, different values of maximum delay and Doppler indices do not have a strong impact on the average coding gain. Meanwhile, it can be observed in the figure that the average coding gain improves with the increase of the squared Euclidean distance ${d_{\rm{E}}^2\left( {\bf{e}} \right)}$.
Furthermore, the derived coding gain bound shows a close match with the overall average coding gain, especially when ${d_{\rm{E}}^2\left( {\bf{e}} \right)}$ is small.

\begin{figure}
\centering
\includegraphics[width=0.6\textwidth]{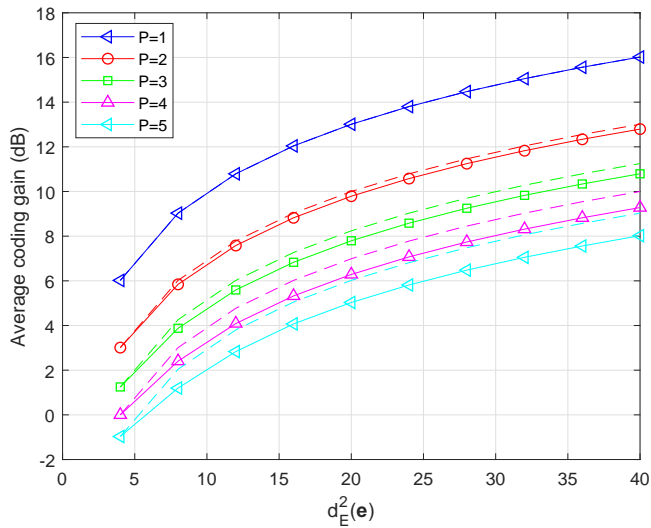}
\caption{Average coding gain for error sequences with ${d_{\rm{E}}^2\left( {\bf{e}} \right)}$ in terms of different numbers of independent resolvable paths, comparing with the coding gain bounds (in dashed lines), where the maximum delay and Doppler indices are set to be $l_{\rm{max}}=2$ and $k_{\rm{max}}=4$, respectively.}
\label{Average_coding_gain_curve2}
\centering
\end{figure}
Fig. \ref{Average_coding_gain_curve2} illustrates the average coding gain and the corresponding coding gain bound (in dashed lines) in decibels (dB) with respect to different values of $P$. Without loss of generality, we set the maximum delay index $l_{\rm{max}}$ and the maximum Doppler index $k_{\rm{max}}$ as $2$ and $4$, respectively.
As shown in the figure, given ${d_{\rm{E}}^2\left( {\bf{e}} \right)}$, the average coding gain decreases with the increase of the number of paths $P$, which is consistent with Corollary 1.
Similar to the previous figure, the coding gain bounds match well with the general behaviour of average coding gains, especially when $P$ is small,
which verifies the correctness of our derivation. On the other hand, we notice that the derived coding gain bound slightly diverge from the average coding gains, when $P$ is large. This observation motivates us to derive a more suitable approximation for the coding gain for large $P$, the details of which will be given in the following subsection.
\subsection{Error Performance Analysis for Large Values of $P$}
When there is a large number of independent resolvable paths of the channel, i.e., the value of $P$ is large, the unconditional PEP can be more accurately bounded by considering the strong law of large number \cite{Yuan2003performance,vucetic2003space}.
In specific, the term $\sum\limits_{i = 1}^P {{\lambda _i}{{| {{{\tilde h}_i}} |}^2}} $ in (\ref{PEP_derivation2}) approaches a Gaussian random variable, due to the central limit theorem \cite{papoulis2002probability}.

Notice that ${{\tilde h}_i}$ follows the complex Gaussian distribution, with mean ${\mu _{{{\tilde h}_i}}}=\mathbb{E}\left[ {\bf{h}} \right]\cdot{{\bf{v}}_i}$ and variance $1/(2P)$ per real dimension.
Therefore, for the ease of derivation, we normalize the variance and rewrite (\ref{PEP_derivation2}) as
\begin{equation}
P\left( {{\left. {{\bf{x}},{\bf{x'}}} \right|{\bf{h}},{{{\bm{\omega}} _\tau }},{{{\bm{\omega}} _\nu }}}} \right) \le \exp \left( { - \frac{{{E_s}}}{{4{N_0}}}\sum\limits_{i = 1}^r {\frac{{{\lambda _i}}}{P}{{\left| {{{\bar h}_i}} \right|}^2}} } \right), \label{New2_PEP_derivation2}
\end{equation}
where ${\bar h_i} = \sqrt P {\tilde h_i}$.
Note that $\left\{ {{{\bar h}_1},{{\bar h}_2},...,{{\bar h}_r}} \right\}$ are independent complex Gaussian random variables with mean ${\mu _{{{\bar h}_i}}}={\sqrt{P}} {\mu _{{{\tilde h}_i}}}$ and variance $1/2$ per real dimension.
Furthermore, it is obvious that $\left| {{{\bar h}_i}} \right|$ follows the Rician distribution with Rician factor ${{\bar{K}}_i}= {\left| {{\mu _{{{\bar h}_i}}}} \right|^2}$ and a unit variance.
Thus, it can be shown that ${{{\left| {{{\bar h}_i}} \right|}^2}}$ follows a noncentral chi-squared distribution with two degrees of freedom (DoFs) and noncentrality parameter $S = {{\bar{K}}_i}$, whose mean and variance are given by
\begin{align}
{\mu _{{{\left| {{{\bar h}_i}} \right|}^2}}} &= 1 + {{\bar{K}}_i}, \label{mean}\\
\sigma _{{{\left| {{{\bar h}_i}} \right|}^2}}^2 &= 1 + 2{{\bar{K}}_i}. \label{variance}
\end{align}
Next, we derive the unconditional PEP by means of Gaussian approximation. To start with, let $\psi  = \sum\limits_{i = 1}^P {{\lambda _i}{{\left| {{{\bar h}_i}} \right|}^2}} $. According to (\ref{mean}) and (\ref{variance}), we approximate $\psi $ as a Gaussian random variable, whose mean is
${\mu _\psi } = \sum\limits_{i = 1}^P {{\lambda _i}\left( {1 + {{{\bar{K}}_i}}} \right)} $ and variance is $\sigma _\psi ^2 = \sum\limits_{i = 1}^P {\lambda _i^2\left( {1 + 2{{\bar{K}}_i}} \right)} $.
Thus, according to the Gaussian distribution of $\psi $, the conditional PEP in (\ref{PEP_derivation2}) is upper-bounded by
\begin{equation}
P\left( {\left. {{\bf{x}},{\bf{x'}}} \right|{{{\bm{\omega}} _\tau }},{{{\bm{\omega}} _\nu }}} \right)\le \int\limits_{0} ^{ + \infty } {\exp \left( { - \frac{{{E_s}}}{{4{N_0}P}}\psi } \right)p} \left( \psi  \right)d\psi  .\label{PEP_derivation6}
\end{equation}
Considering
\begin{align}
\int\limits_0^{ + \infty } {\exp \left( { - \gamma \psi } \right)p} \left( \psi  \right)d\psi
 = \exp \left( {\frac{1}{2}{\gamma ^2}\sigma _\psi ^2 - \gamma {\mu _\psi }} \right)Q\left( {\frac{{\gamma \sigma _\psi ^2 - {\mu _\psi }}}{{{\sigma _\psi }}}} \right),\gamma  > 0,
\end{align}
we obtain
\begin{align}
P\left( {\left. {{\bf{x}},{\bf{x'}}} \right|{{{\bm{\omega}} _\tau }},{{{\bm{\omega}} _\nu }}} \right) \le \exp \left( {\frac{1}{2}{{\left( {\frac{{{E_s}}}{{4{N_0}}}} \right)}^2}\cdot\frac{{\sigma _\psi ^2}}{{{P^2}}} - \frac{{{E_s}}}{{4{N_0}}}\cdot\frac{{{\mu _\psi }}}{P}} \right)
Q\left( {\frac{{{E_s}}}{{4{N_0}}}\cdot\frac{{{\sigma _\psi }}}{P} - \frac{{{\mu _\psi }}}{{{\sigma _\psi }}}} \right). \label{PEP_derivation7}
\end{align}
Similar to the previous subsection, we consider the special case of Rayleigh fading.

In the case of \textbf{Rayleigh fading}, i.e., $|{{{\bar{h}}_i}}|$ and $|{{h_i}}|$ follow the Rayleigh distribution, we have ${\mu _\psi } = \sum\limits_{i = 1}^P {{\lambda _i}} $ and $\sigma _\psi ^2 = \sum\limits_{i = 1}^P {\lambda _i^2} $.
Therefore, the right hand side of (\ref{PEP_derivation7}) is given by
\begin{align}
P\left( {\left. {{\bf{x}},{\bf{x'}}} \right|{{{\bm{\omega}} _\tau }},{{{\bm{\omega}} _\nu }}} \right) \le& \exp \left( {\frac{1}{2}{{\left( {\frac{{{E_s}}}{{4{N_0}}}} \right)}^2}\sum\limits_{i = 1}^P {\frac{{\lambda _i^2}}{{{P^2}}}}  - \frac{{{E_s}}}{{4{N_0}}}\sum\limits_{i = 1}^P {\frac{{{\lambda _i}}}{P}} } \right)\notag\\
&\quad\quad Q\left( {\frac{{{E_s}}}{{4{N_0}P}}\sqrt {\sum\limits_{i = 1}^P {\lambda _i^2} }  - \frac{{\sum\nolimits_{i = 1}^P {{\lambda _i}} }}{{\sqrt {\sum\nolimits_{i = 1}^P {\lambda _i^2} } }}} \right).
\label{PEP_derivation8}
\end{align}
Furthermore, we consider the Chernoff bound of the Q-function \cite{Yuan2003performance}
\begin{equation}
Q\left( \gamma  \right) \le \exp \left( { - \frac{1}{2}{\gamma ^2}} \right),\gamma  > 0,
\end{equation}
and (\ref{PEP_derivation8}) can be further upper-bounded by
\begin{align}
P\left( {\left. {{\bf{x}},{\bf{x'}}} \right|{{{\bm{\omega}} _\tau }},{{{\bm{\omega}} _\nu }}} \right) \le &\exp \left( {\frac{1}{2}{{\left( {\frac{{{E_s}}}{{4{N_0}}}} \right)}^2}\sum\limits_{i = 1}^P {\frac{{\lambda _i^2}}{{{P^2}}}}  - \frac{{{E_s}}}{{4{N_0}}}\sum\limits_{i = 1}^P {\frac{{{\lambda _i}}}{P}} } \right)\notag\\
&\exp \left( { - \frac{1}{2}{{\left( {\frac{{{E_s}}}{{4{N_0}}}} \right)}^2}\sum\limits_{i = 1}^P {\frac{{\lambda _i^2}}{{{P^2}}}}  - \frac{{{{\left( {\sum\limits_{i = 1}^P {{\lambda _i}} } \right)}^2}}}{{2\sum\limits_{i = 1}^P {\lambda _i^2} }} + \frac{{{E_s}}}{{4{N_0}}}\sum\limits_{i = 1}^P {\frac{{{\lambda _i}}}{P}} } \right) \notag\\
=&\exp \left( { - \frac{{{{\left( {\sum\limits_{i = 1}^P {{\lambda _i}} } \right)}^2}}}{{2\sum\limits_{i = 1}^P {\lambda _i^2} }}} \right),
\label{New_PEP_Large_P_derivation2}
\end{align}
when
\begin{equation}
\frac{{{E_s}}}{{4{N_0}}} \ge \frac{P{\sum\nolimits_{i = 1}^P {{\lambda _i}} }}{{\sum\nolimits_{i = 1}^P {\lambda _i^2} }}. \label{New_PEP_Large_P_derivation1}
\end{equation}
Based on (\ref{New_PEP_Large_P_derivation2}), the unconditional PEP can be approximately upper-bounded as shown in the following Theorem.

\textbf{Theorem 3} (\emph{Unconditional PEP upper bound for large $P$}):
For a large value of $P$ and a reasonably high SNR, i.e., $\frac{{{E_s}}}{{4{N_0}}} \ge \frac{P}{{d_{\rm{E}}^2\left( {\bf{e}} \right)}}$, the unconditional PEP of OTFS systems can be approximately upper-bounded by
\begin{equation}
P\left(  {{\bf{x}},{\bf{x'}}} \right) \mathbin{\lower.3ex\hbox{$\buildrel<\over
{\smash{\scriptstyle\sim}\vphantom{_x}}$}} \exp \left( { - \frac{{{E_s}}}{{8{N_0}}}d_{\rm{E}}^2} \left( {\bf{e}} \right)\right) . \label{Large_P_upper_bound}
\end{equation}

\emph{Proof}: The proof is given in Appendix D.

It should be noted that, for $P \ge 4$, the approximation in Theorem 3 is sufficiently accurate owing to the strong law of the large number \cite{vucetic2003space,Yuan2003performance}.
On the other hand, the number of paths $P$ is usually smaller than the squared Euclidean distance $d_{\rm{E}}^2\left( {\bf{e}} \right)$ for practical wireless transmissions with good channel codes \footnote{For example, a popular industry-standard rate-$1/2$ convolutional code with code memory of $6$ has a minimum squared Euclidean distance $d_{\rm{E}}^2 \left( {\bf{e}} \right)=40$ \cite{ryan2009channel}.}.
Therefore, our SNR assumption is reasonable.
Compared with Corollary 1, it is not surprising that the unconditional PEP only depends on the squared Euclidean distance $d_{\rm{E}}^2\left( {\bf{e}} \right)$, regardless of the delay and Doppler indices.
Furthermore, it should be noted that (\ref{Large_P_upper_bound}) is of the similar form of the error performance for AWGN channels \cite{Ventura1997impact}.
This is because the impact of fading is mitigated by a large number of diversity branches and consequently, in order words, the channel with a large number of diversity paths approaches
an AWGN model \cite{Ventura1997impact}.

In this section, we have derived the error performance analysis of coded OTFS systems. It should be noted that our error performance analysis only considers the integer Doppler case. However, the extension of the above analysis to the fractional Doppler case is straightforward.
In the following, we will design suitable channel codes according to our error performance analysis.

\subsection{Code Design Issues}
According to the derived analysis, the rule-of-thumb channel code design criterion is discussed in this section.
Without loss of generality, we only consider the Rayleigh fading channel in the following.
It should be noticed from the previous analysis that the code design criterion for the coded OTFS system is to maximize the minimum squared Euclidean distance $d_{\rm{E}}^2\left( {\bf{e}} \right)$.

\textbf{Proposition 1} \emph{(The squared Euclidean distance criterion)}:
The channel code should be designed to maximize the minimum squared Euclidean distance among all pairs of possible codewords.


We note that even with the designed code, the error performance of coded OTFS systems may still vary with different channel parameters e.g., ${\bm{\omega}} _\tau$ and ${\bm{\omega}} _\nu$.
This detrimental effect due to channel realizations is widely observed in the system designs for fading channels, such as in \cite{Lu2000space,Biglieri1998fading}.
In specific, with different channel parameters, the value of the conditional coding gain can be different even with the same error sequence, which may potentially jeopardize the overall error performance of OTFS systems. In order to obtain a more robust performance, it is desirable to apply an interleaver to permute the coded symbols before sending to the constellation mapper or the OTFS modulator in the DD domain.
As pointed out in \cite{Biglieri1998fading}, such an interleaver can ``whiten" the transmitted symbols from the information theoretic point of view and the detrimental effect on the error performance due to the channel parameters can thus be alleviated.

To examine our performance analysis of the coded OTFS systems, we perform numerical simulations for OTFS systems over high-mobility channels, the results of which will be shown in the next section.

\section{Numerical Results}
In this section, the error performance of the coded OTFS system with various channel codes is evaluated via numerical simulations.
We consider the BPSK signal for the OTFS system, where the data sequence is firstly encoded and interleaved, and then BPSK mapped, according to Fig. {\ref{System_Model}}.
Without loss of generality, we consider the maximum-likelihood (ML) detection of OTFS modulation. In order to verify the accuracy of the analytical results, we consider four different convolutional codes (with trellis termination) with different minimum squared Euclidean distance $d_{\rm{E}}^2{\left( {\bf{e}} \right)} $ among all possible codeword pairs. The details of the code parameters are given in Table \ref{Code_parameters}, including the generator matrix and the memory length. In particular, we also show the smallest value of squared Euclidean distance $d_{\rm{E}}^2\left( {\bf{e}} \right)$ among all possible codeword pairs for each code. In specific, we consider a coded OTFS system with $N=8$ and $M=16$ and correspondingly the codeword length for all considered simulations is $128$ bits unless otherwise specified.
The channel decoder adopts the logarithm domain Bahl-Cocke-Jelinek-Raviv (BCJR) algorithm \cite{Bahl1974optimal}.
Furthermore, we consider the Rayleigh fading case.
If not otherwise specified, we set the maximum delay index as $l_{\max }=3$ and the maximum Doppler index as $k_{\max }=5$,
which is corresponding to a relative UE speed around $250$ km/h with $4$ GHz carrier frequency and $1.5$ kHz sub-carrier spacing. For each channel realization, we randomly select the
delay and Doppler indices such that we have $ - {k_{\max }} \le {k_i} \le {k_{\max }}$ and $0 \le {l_i} \le {l_{\max }}$.


\begin{table*}[htbp]
\vspace{-3mm}
\caption{Code Parameters}
\centering
\begin{tabular}{|c|c|c|c|}
\hline
Code structure~&~Generator matrix~&~Memory length~&~Minimum $d_{\rm{E}}^2\left( {\bf{e}} \right)$\\
\hline
A~&~$\left[ 1+D, D \right]$&~1~&~12\\
\hline
B~&~$\left[ 1 + {D^2}, 1 + D + {D^2} \right]$&~2~&~20\\
\hline
C~&~$\left[ {1 + {D^2} + {D^5},1 + D + {D^2} + {D^3} + {D^4} + {D^5}} \right]$~&~5~&~32\\
\hline
D~&~$\left[ {1 + D + {D^2} + {D^5} + {D^6},1 + {D^2} + {D^3} + {D^4} + {D^6}} \right]$&~6~&~40 \\
\hline
\end{tabular}
\label{Code_parameters}
\end{table*}
\begin{figure}
\centering
\includegraphics[width=0.6\textwidth]{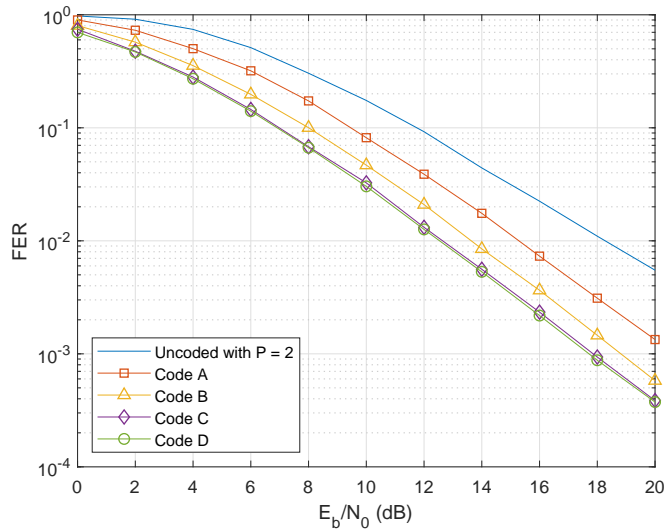}
\caption{FER performance for OTFS modulation with different codes for $P=2$, where the relative UE speed is $250$ km/h.}
\label{Path2}
\centering
\vspace{-3mm}
\end{figure}
The frame-error-rate (FER) performance of the OTFS systems with $P=2$ is shown in Fig. \ref{Path2}.
It can be observed that the slope of the FER curve for the uncoded OTFS system is slightly worse than that for coded OTFS systems. This indicates that the uncoded OTFS system with $P=2$ does not guarantee the full diversity for all possible channel realizations, which is consistent with the analysis in \cite{Raviteja2019effective}.
More importantly, this observation also shows that the application of channel coding can improve the diversity gain of OTFS systems in the case where the OTFS modulation fails to achieve the full diversity. 
Moreover, we observe that employing the channel code with a larger minimum squared Euclidean distance $d_{\rm{E}}^2\left( {\bf{e}} \right)$ indeed leads to a larger coding gain. In specific, we observe from the figure that for FER $\approx {10^{ - 2}}$, the required SNRs for codes A, B, C and D, are $15.28$ dB, $13.64$ dB, $12.65$ dB, and $12.54$ dB, respectively. Compared to uncoded OTFS systems, these four coded OTFS systems achieve coding gains roughly $2.99$ dB, $4.63$ dB, $5.62$ dB, and $5.73$ dB, respectively. This observation clearly substantiates the proposed performance analysis and code design criterion in Proposition 1.


\begin{figure}
\centering
\includegraphics[width=0.6\textwidth]{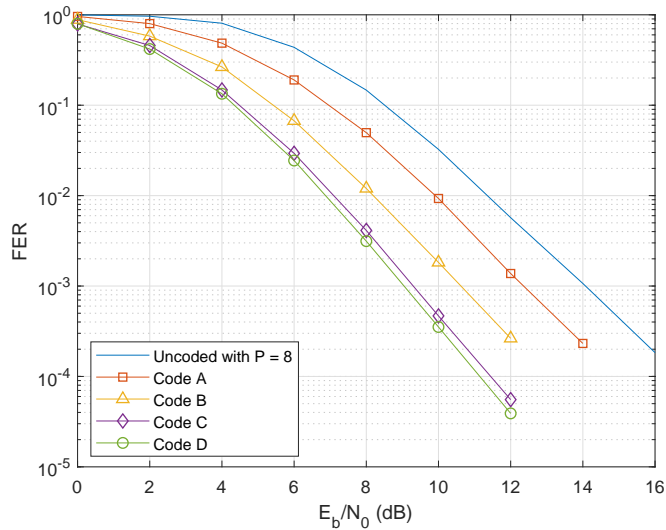}
\caption{FER performance for OTFS modulation with different codes for $P=8$, where the relative UE speed is $250$ km/h.}
\label{Path8}
\centering
\vspace{-3mm}
\end{figure}
Fig. \ref{Path8} shows the FER performance of the OTFS systems with $P=8$.
We notice that the slope of the FER curve for the uncoded OTFS system with $P=8$ is slightly lower than that for coded OTFS systems, which is consistent with the previous figure.
On the other hand, similar to Fig. \ref{Path2}, we observe that the channel code with a larger minimum squared Euclidean distance $d_{\rm{E}}^2\left( {\bf{e}} \right)$ enjoys a larger coding gain compared to the uncoded OTFS system, which is consistent with the analysis of Theorem 3, i.e., the channel approaches
an AWGN model with a large number of diversity branches.
Together with the observations from Fig. \ref{Path2}, one can conclude that our proposed code design criterion in Proposition 1 is universal for general OTFS systems in spite of the channel parameters and the number of paths $P$.

\begin{figure}
\centering
\includegraphics[width=0.6\textwidth]{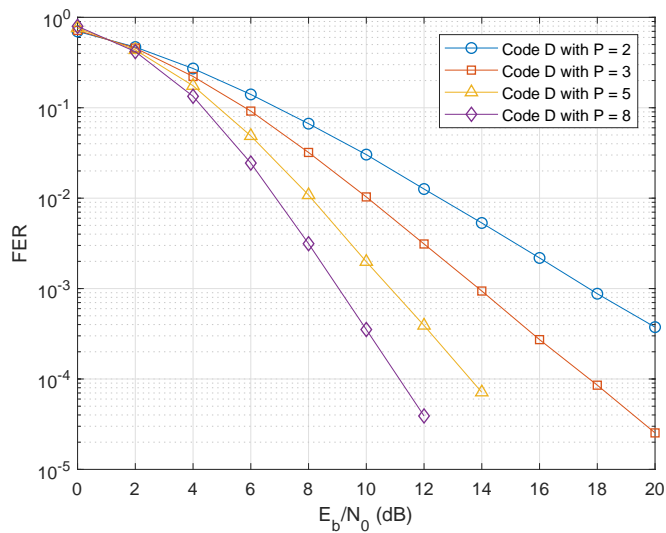}
\caption{FER performance of code D for OTFS modulation with different number of paths and the relative UE speed $250$ km/h.}
\label{Same_code_dif_path}
\centering
\vspace{-5mm}
\end{figure}
The FER performance of the OTFS modulation with code D and different number of paths $P$ is illustrated in Fig. \ref{Same_code_dif_path}.
It can be observed from the figure that given the same code, the error performance of the coded OTFS systems improves with the increase of number of distinguishable paths. Furthermore, it is obvious that
with the same code, a larger value of $P$ corresponds to a larger diversity advantage as indicated in the figure, which is consistent with our analysis.

\begin{figure}
\centering
\includegraphics[width=0.6\textwidth]{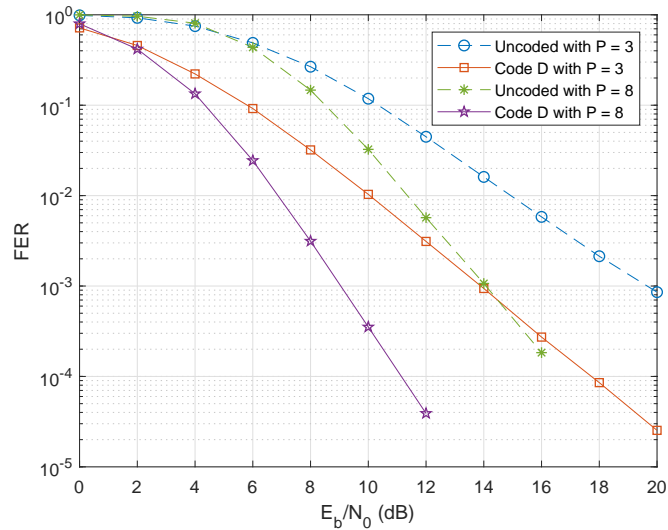}
\caption{FER performance of code D for OTFS modulation with $P=3$ and $P=8$, comparing with that of uncoded OTFS systems, where the relative UE speed is $250$ km/h.}
\label{Coding_vs_Path}
\centering
\vspace{-3mm}
\end{figure}

Fig. \ref{Coding_vs_Path} presents the trade-off between the diversity and coding gain. In particular, we consider the FER performance of code D with $P = 3$ and $P = 8$, comparing with that of the corresponding uncoded OTFS systems.
As shown in figure, at FER $\approx {10^{ - 3}}$, the coded OTFS system with $P = 3$ exhibits around $5.7$ dB coding gain compared to that of the uncoded OTFS system with the same FER,
while only around $5.0$ dB coding gain is obtained for the coded OTFS system with $P = 8$. This observation matches the prediction in Proposition 1, which
indicates that the coding gain reduces with the increase of $P$, given the same channel code.

\begin{figure}
\centering
\includegraphics[width=0.6\textwidth]{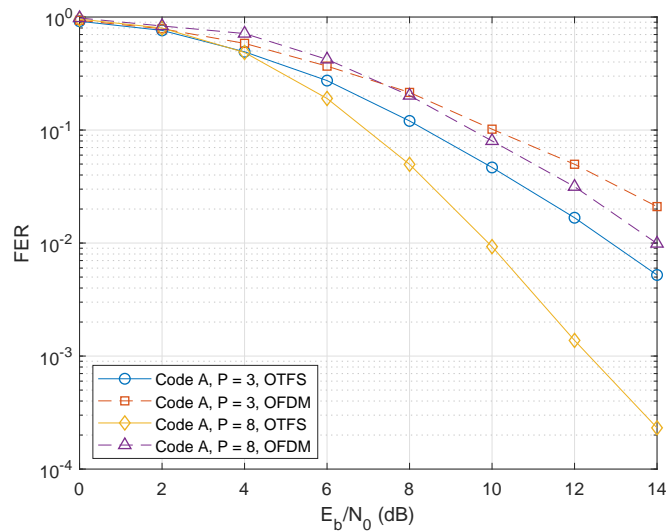}
\caption{FER performance of code A for OTFS modulation with $P=3$ and $P=8$, compared to an OFDM system, where the relative UE speed is $250$ km/h.}
\label{OTFS_vs_OFDM}
\centering
\vspace{-3mm}
\end{figure}

We compare the FER performance of code A with $P = 3$ and $P = 8$, and that of the corresponding OFDM systems in Fig. \ref{OTFS_vs_OFDM}. As observed from the figure, the OTFS system enjoy better error performance than that of the OFDM system with the same code, for both $P = 3$ and $P = 8$. Furthermore, the FER curve of the OFDM system shares almost the same slope as that of the OTFS system, for $P = 3$. As for $P = 8$, the achieved diversity gain of the OTFS system is clearly higher than that of the OFDM system.
Note that the diversity gain of coded OFDM systems is determined by the smaller value of the minimum symbol-wise Hamming distance of the code ${\delta _{\rm{H}}}$ and the number of paths $P$\footnote{According to \cite{Diversity_OFDM}, we have $r_{\rm{OFDM}}={\rm{min}}({\delta _{\rm{H}}},P)$, where $r_{\rm{OFDM}}$ is the achievable diversity gain of a coded OFDM system. In specific, we have $r_{\rm{OFDM}}=3$ for both $P=3$ and $P=8$, with Code A.} \cite{Diversity_OFDM}, while OTFS systems can obtain the full diversity almost surely regardless of the employment of channel codes.
Therefore, this observation clearly shows the advantage of the OTFS systems over the OFDM systems.

\begin{figure}
\centering
\includegraphics[width=0.6\textwidth]{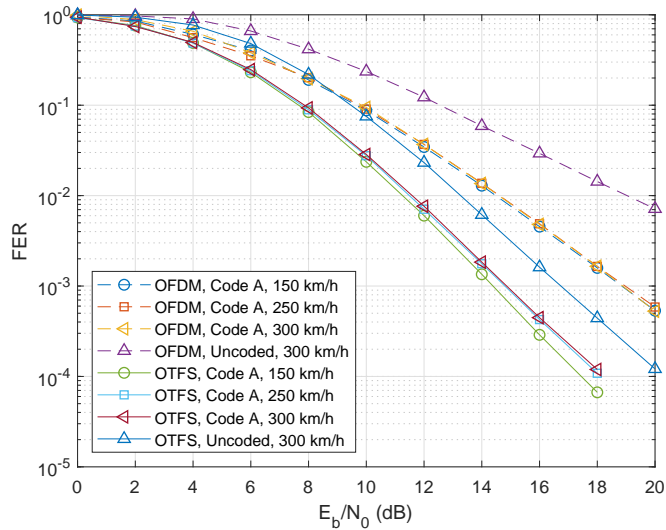}
\caption{FER performance of code A for OTFS modulation and OFDM modulation with $P=4$, where the values of the relative UE speed are $150$ km/h, $250$ km/h, and $300$ km/h, respectively.}
\label{OTFS_vs_OFDM_vs_Speed}
\centering
\vspace{-3mm}
\end{figure}

The FER performance of both OTFS systems and OFDM systems with various values of the relative UE speed is compared in Fig. \ref{OTFS_vs_OFDM_vs_Speed}. We consider $l_{\max }=3$ and $k_{\max }=3, 5, 6$, which corresponds to the cases where the relative UE speeds are $150$ km/h, $250$ km/h, and $300$ km/h, respectively.
We apply the ML detection for both the OTFS and OFDM systems to have a fair comparison. We observe that the error performance of both OTFS and OFDM systems with ML detection does not change much with different relative UE speeds, which is consistent with the observations in~\cite{Raviteja2018interference}.
Note that, the ML detection is not practically feasible for conventional OFDM systems due to the high detection complexity. Therefore, frequency domain equalization is usually deployed for OFDM systems~\cite{mostofi2005ici}. In this case, the error performance of OFDM systems will degrade dramatically with the increase of the speed~\cite{mostofi2005ici,Yuan2020Orthogonal_report} due to the severe inter-carrier interference (ICI) induced by the Doppler spread.
Furthermore, the FER performance of the OTFS systems outperform that of the OFDM systems, including both coded and uncoded cases with various values of the speed.
Similar to the previous figure, we also observe that the achieved diversity gain of the OTFS system is higher than that of the OFDM system, which agrees with our analysis.

\begin{table*}[htbp]
\vspace{-3mm}
\caption{Code Parameters for Fig. \ref{LDPC_Turbo}}
\centering
\begin{tabular}{|c|c|c|c|}
\hline
Code~&~Data length~&~Codeword length~&~Code rate\\
\hline
Convolutional Code D~&~250&~512~&~0.488\\
\hline
5G LDPC~&~256&~512~&~0.5\\
\hline
LTE Turbo~&~250~&~512~&~0.488\\
\hline
\end{tabular}
\label{Code_parameters2}
\end{table*}

\begin{figure}
\centering
\includegraphics[width=0.6\textwidth]{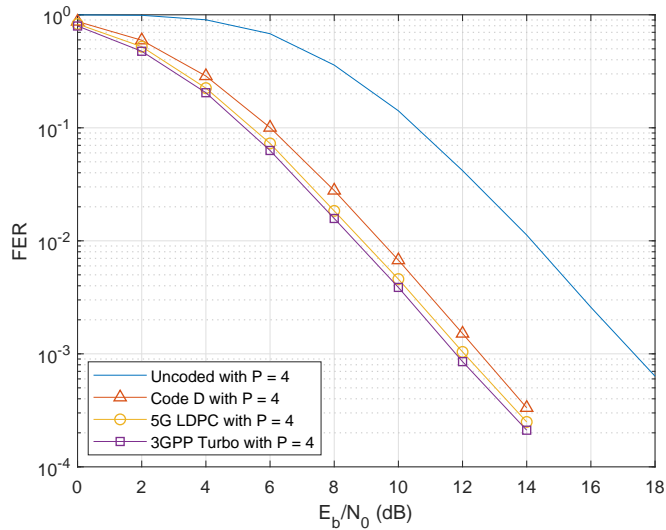}
\caption{FER performance of 5G LDPC code and 3GPP Turbo code for OTFS modulation with $P=4$ and a relative UE speed $250$ km/h, where the codeword length is $512$ bits.}
\label{LDPC_Turbo}
\centering
\vspace{-4mm}
\end{figure}

We present the FER results of coded OTFS systems with modern codes in Fig. \ref{LDPC_Turbo}. We consider an LDPC code from the 5G communication standard \cite{5gChannel1} (referred to as the 5G LDPC code) and the Turbo code from the 3GPP long term evolution (LTE) standard \cite{LTE_Turbo_TS} (referred to as the LTE Turbo code), where code parameters are given in Table \ref{Code_parameters2} and we have $N=16$ and $M=32$ for OTFS modulation.
As a benchmark, the FER performance of the convolutional code D is also given in Fig. \ref{LDPC_Turbo}. We observe that, the LTE Turbo code achieves the best error performance compared to the 5G LDPC code and the convolutional code D with $P=4$, although they all share the same diversity gain. More specifically, at FER $\approx {10^{ - 3}}$, 5G LDPC code and LTE Turbo code show around $0.5$ dB and $0.7$ dB SNR gain compared to the convolutional code D. Note that a similar observation of the error performance can be observed over the AWGN channel, where the LTE Turbo code has the best performance while the convolutional code D has the worst performance. Therefore, this observation indicate that codes optimized for AWGN channels can also achieve a good error performance in the OTFS systems, which is consistent with our analysis.

\section{Conclusion}
In this paper, we studied the performance analysis of coded OTFS systems over high-mobility channels. We first derived the conditional PEP for a given channel realization and then obtained the unconditional PEP by leveraging proper bounding techniques. We discussed two cases of the OTFS transmission according to the number of independent resolvable paths of the channel, where we showed that the coding improvement of OTFS systems depends on the squared Euclidean distance between a pair of codewords. More importantly, we demonstrated the fundamental trade-off between the diversity gain and the coding gain for OTFS systems.
Furthermore, we proposed a code design criterion based on the derived unconditional bound.
The analysis and the code design criterion are verified by numerical simulations with various channel codes.
Our future work will focus on the modern code design for OTFS systems by considering analytical tools such as the density evolution and the extrinsic information transfer (EXIT) chart.

\appendices
\section{Proof of Lemma 3}
Notice that, ${{\bf{\Omega }}\left( {\bf{e}} \right)}$ is positive definite Hermitian. Hence, the eigenvalues $\left\{ {{\lambda _i}} \right\}$ of ${{\bf{\Omega }}\left( {\bf{e}} \right)}$ are all positive.
Considering the arithmetic mean and geometric mean (AM-GM) inequality, we obtain
\begin{equation}
\sum\limits_{i = 1}^P {\frac{1}{{{\lambda _i}}}}   \ge P{\left( {\prod\limits_{i = 1}^P {\frac{1}{{{\lambda _i}}}} } \right)^{\frac{1}{P}}} = \frac{P}{{{{\left( {\prod\limits_{i = 1}^P {{\lambda _i}} } \right)}^{\frac{1}{P}}}}}.
\label{Lm1_1}
\end{equation}
Then, we apply the Cauchy-Schwarz inequality to the denominator of the right hand side of (\ref{Lm1_1}), which yields
\begin{equation}
\frac{P}{{{{\left( {\prod\limits_{i = 1}^P {{\lambda _i}} } \right)}^{\frac{1}{P}}}}} \ge \frac{{{P^2}}}{{\sum\limits_{i = 1}^P {{\lambda _i}} }}\ge \frac{P}{{d_{\rm{E}}^2\left( {\bf{e}} \right)}}.
\end{equation}
It is obvious that the equality only holds when the eigenvalues $\left\{ {{\lambda _i}} \right\}$ are of the same value, e.g., the codeword difference matrix ${{\bf{\Omega }}\left( {\bf{e}} \right)}$ is a diagonal matrix, i.e., ${\bf{\Omega }}\left( {\bf{e}} \right)=\textrm{diag}\left\{ {d_{\rm{E}}^2{\left( {\bf{e}} \right)}, \ldots ,d_{\rm{E}}^2}{\left( {\bf{e}} \right)} \right\}$. This completes the proof of Lemma 3.

\section{Proof of Theorem 1}
The determinant of the codeword difference matrix ${{\bf{\Omega }}\left( {\bf{e}} \right)}$ can be written as
\begin{equation}
\det \left( {{\bf{\Omega }}\left( {\bf{e}} \right)} \right) = \exp \left( {\ln \left( {\prod\limits_{i = 1}^P {{\lambda _i}} } \right)} \right) = \exp \left( {\sum\limits_{i = 1}^P {\ln \left( {{\lambda _i}} \right)} } \right),
\end{equation}
where $\{\lambda _i\}$ for $i \in \left\{ {1, \ldots ,P} \right\}$ are the eigenvalues of ${{\bf{\Omega }}\left( {\bf{e}} \right)}$.
Furthermore, let us consider the inequality $\ln \left( \gamma \right) \ge  1 - \frac{1}{\gamma },  \gamma \in \left( {0, + \infty } \right)$,
where the equality only holds when $\gamma=1$.
Therefore, we have
\begin{align}
\det \left( {{\bf{\Omega }}\left( {\bf{e}} \right)} \right) &= \exp \left( {\sum\limits_{i = 1}^P {\ln \left( {{\lambda _i}} \right)} } \right) \notag\\
&= \exp \left( {P\ln \left( {d_{\rm{E}}^2\left( {\bf{e}} \right)} \right) + \sum\limits_{i = 1}^P {\ln \left( {\frac{{{\lambda _i}}}{{d_{\rm{E}}^2\left( {\bf{e}} \right)}}} \right)} } \right)\notag\\
& \ge  \exp \left( {P\ln \left( {d_{\rm{E}}^2\left( {\bf{e}} \right)} \right) + \sum\limits_{i = 1}^P {\left( {1 - \frac{{d_{\rm{E}}^2\left( {\bf{e}} \right)}}{{{\lambda _i}}}} \right)} } \right) \label{inequality}\\
&={\left( {d_{\rm{E}}^2\left( {\bf{e}} \right)} \right)^P}\exp \left( {P - d_{\rm{E}}^2\left( {\bf{e}} \right){\rm{tr}}\left( {{{\left( {{\bf{\Omega }}\left( {\bf{e}} \right)} \right)}^{ - 1}}} \right)} \right).
\end{align}
It is obvious that the equality holds in (\ref{inequality}) only if all the eigenvalues $\{\lambda _i\}$ of ${{\bf{\Omega }}\left( {\bf{e}} \right)}$ equal to ${d_{\rm{E}}^2\left( {\bf{e}} \right)}$.
Notice that the eigenvalues $\{\lambda _i\}$ of ${{\bf{\Omega }}\left( {\bf{e}} \right)}$ equal to the main diagonal elements when ${{\bf{\Omega }}\left( {\bf{e}} \right)}$ is a diagonal matrix, i.e., ${\bf{\Omega }}\left( {\bf{e}} \right)=\textrm{diag}\left\{ {d_{\rm{E}}^2\left( {\bf{e}} \right), \ldots ,d_{\rm{E}}^2\left( {\bf{e}} \right)} \right\}$, in which case we have $\det \left( {{\bf{\Omega }}\left( {\bf{e}} \right)} \right) = {\left( {d_{\rm{E}}^2}\left( {\bf{e}} \right) \right)^P}$.
This completes the proof of Theorem 1.

\section{Proof of Theorem 2}
Following Theorem 1, we note that the term $ {P - {d_{\rm{E}}^2\left( {\bf{e}} \right)}{\rm{tr}}\left( {{\bf{\Omega }}_i^{ - 1}} \right)}$ is less than zero according to Lemma 3.
Therefore, the value of $\exp \left( {P - {d_{\rm{E}}^2\left( {\bf{e}} \right)}{\rm{tr}}\left( {{\bf{\Omega }}_i^{ - 1}} \right)} \right)$ changes slowly with the increase of ${\rm{tr}}\left( {{\bf{\Omega }}_i^{ - 1}} \right)$, owing to the slow decay property of the corresponding function.
Therefore, we apply the result from Lemma 3 to approximate the determinant of ${{\bf{\Omega }}\left( {\bf{e}} \right)}$ \cite{Yuan2003performance,vucetic2003space},
\begin{align}
\det \left( {{\bf{\Omega }}\left( {\bf{e}} \right)} \right) &\ge \exp \left( {P - d_{\rm{E}}^2\left( {\bf{e}} \right){\rm{tr}}\left( {{\bf{\Omega }}_i^{ - 1}} \right)} \right) \notag\\
&\mathbin{\lower.3ex\hbox{$\buildrel>\over
{\smash{\scriptstyle\sim}\vphantom{_x}}$}}\exp \left( {P - d_{\rm{E}}^2\left( {\bf{e}} \right){{\left( {{\rm{tr}}\left( {{\bf{\Omega }}_i^{ - 1}} \right)} \right)}_{\min }}} \right)\notag\\
&={{{\left( {d_{\rm{E}}^2\left( {\bf{e}} \right)} \right)}^P}\exp \left( {P - d_{\rm{E}}^2\left( {\bf{e}} \right)\frac{P}{{d_{\rm{E}}^2\left( {\bf{e}} \right)}}} \right)}\notag\\
&={{{\left( {d_{\rm{E}}^2\left( {\bf{e}} \right)} \right)}^P}},
\end{align}
where the equality holds if ${\bf{\Omega }}\left( {\bf{e}} \right)$ is a diagonal matrix, i.e., ${\bf{\Omega }}\left( {\bf{e}} \right)=\textrm{diag}\left\{ {d_{\rm{E}}^2{\left( {\bf{e}} \right)}, \ldots ,d_{\rm{E}}^2}{\left( {\bf{e}} \right)} \right\}$.
Mathematically, the above approximation may be loose if ${\bf{\Omega }}\left( {\bf{e}} \right)$ is ill-conditioned. Therefore, we justify the accuracy of our approximation as follows.
A commonly adopted approach in testifying if a matrix is in ill-condition is the $\bf{P}$-condition number \cite{taylor1978condition}. In specific, we consider the lower bound on the $\bf{P}$-condition number of a Gram matrix \cite{taylor1978condition}. The ${\bf{P}}$-condition number of ${\bf{\Omega }}\left( {\bf{e}} \right)$ is defined by
\begin{equation}
{\bf{P}}\left( {\bf{\Omega }}\left( {\bf{e}} \right)\right) \buildrel \Delta \over = r\left( {\bf{\Omega }}\left( {\bf{e}} \right) \right)r\left( {\left( {{\bf{\Omega }}\left( {\bf{e}} \right)} \right)^{ - 1}} \right),
\end{equation}
where $r\left( {{\bf{\Omega }}\left( {\bf{e}} \right)} \right)$ is the \emph{spectral radius} of ${{\bf{\Omega }}\left( {\bf{e}} \right)}$, i.e., $r\left( {{\bf{\Omega }}\left( {\bf{e}} \right)} \right)$ equals to the largest eigenvalues of ${{\bf{\Omega }}\left( {\bf{e}} \right)}$.
In particular, the matrix ${{\bf{\Omega }}\left( {\bf{e}} \right)}$ is said to be ill-conditioned if ${\bf{P}}\left({{\bf{\Omega }}\left( {\bf{e}} \right)} \right) $ is large and is to be well-conditioned if ${\bf{P}}\left({{\bf{\Omega }}\left( {\bf{e}} \right)} \right) $ is small.
According to \cite{taylor1978condition}, we have ${\bf{P}}\left( {{\bf{\Omega }}\left( {\bf{e}} \right)} \right)\ge {{{{\left( {{{\left\| {{{\bf{u}}_i}} \right\|}^2}} \right)}_{\max }}} \mathord{\left/
 {\vphantom {{{{\left( {{{\left\| {{{\bf{u}}_i}} \right\|}^2}} \right)}_{\max }}} {{{\left( {{{\left\| {{{\bf{u}}_j}} \right\|}^2}} \right)}_{\min }}}}} \right.
 \kern-\nulldelimiterspace} {{{\left( {{{\left\| {{{\bf{u}}_j}} \right\|}^2}} \right)}_{\min }}}}$ for $1 \le i,j \le P$, which yields
\begin{equation}
{\bf{P}}\left( {{\bf{\Omega }}\left( {\bf{e}} \right)} \right)\ge  {{d_{\rm{E}}^2\left( {\bf{e}} \right)} \mathord{\left/
 {\vphantom {{d_{\rm{E}}^2\left( {\bf{e}} \right)} {d_{\rm{E}}^2\left( {\bf{e}} \right) = 1}}} \right.
 \kern-\nulldelimiterspace} {d_{\rm{E}}^2\left( {\bf{e}} \right) = 1}}.
\end{equation}
We can see that the $\bf{P}$-condition number of ${{\bf{\Omega }}\left( {\bf{e}} \right)}$ always greater than or equal to $1$, which indicates that ${{\bf{\Omega }}\left( {\bf{e}} \right)}$ is generally well-conditioned.
This completes the proof of Theorem 2.

\section{Proof of Theorem 3}

Recalling (\ref{New_PEP_Large_P_derivation1}), we note that (\ref{New_PEP_Large_P_derivation2}) only holds if
\begin{equation}
\sum\nolimits_{i = 1}^P {\lambda _{i}^2}  \ge \frac{{P\sum\nolimits_{i = 1}^P {{\lambda _{i}}} }}{{{{{E_s}} \mathord{\left/
 {\vphantom {{{E_s}} {\left( {4{N_0}} \right)}}} \right.
 \kern-\nulldelimiterspace} {\left( {4{N_0}} \right)}}}}=\frac{{{P^2}d_{\rm{E}}^2\left( {\bf{e}} \right)}}{{{{{E_s}} \mathord{\left/
 {\vphantom {{{E_s}} {\left( {4{N_0}} \right)}}} \right.
 \kern-\nulldelimiterspace} {\left( {4{N_0}} \right)}}}}. \label{Th4_derivation1}
\end{equation}
Therefore, we consider the approximation of (\ref{New_PEP_Large_P_derivation2}) as follows
\begin{align}
P\left(  {{\bf{x}},{\bf{x'}}} \right)&\le \exp \left( { - {{{{\left( {\sum\limits_{i = 1}^P {{\lambda _i}} } \right)}^2}} \mathord{\left/
 {\vphantom {{{{\left( {\sum\limits_{i = 1}^P {{\lambda _i}} } \right)}^2}} {\left( {2\sum\limits_{i = 1}^P {\lambda _i^2} } \right)}}} \right.
 \kern-\nulldelimiterspace} {\left( {2\sum\limits_{i = 1}^P {\lambda _i^2} } \right)}}} \right)
  \mathbin{\lower.3ex\hbox{$\buildrel<\over
{\smash{\scriptstyle\sim}\vphantom{_x}}$}} \exp \left( { - {{{{\left( {\sum\limits_{i = 1}^P {{\lambda _i}} } \right)}^2}} \mathord{\left/
 {\vphantom {{{{\left( {\sum\limits_{i = 1}^P {{\lambda _i}} } \right)}^2}} {2{{\left( {\sum\limits_{i = 1}^P {\lambda _i^2} } \right)}_{\min }}}}} \right.
 \kern-\nulldelimiterspace} {2{{\left( {\sum\limits_{i = 1}^P {\lambda _i^2} } \right)}_{\min }}}}} \right) \notag\\
& =\exp \left( { - {{{{\left( {Pd_{\rm{E}}^2\left( {\bf{e}} \right)} \right)}^2}} \mathord{\left/
 {\vphantom {{{{\left( {Pd_{\rm{E}}^2\left( {\bf{e}} \right)} \right)}^2}} {\left( {2\frac{{{P^2}d_{\rm{E}}^2\left( {\bf{e}} \right)}}{{{{{E_s}} \mathord{\left/
 {\vphantom {{{E_s}} {\left( {4{N_0}} \right)}}} \right.
 \kern-\nulldelimiterspace} {\left( {4{N_0}} \right)}}}}} \right)}}} \right.
 \kern-\nulldelimiterspace} {\left( {\frac{{2{P^2}d_{\rm{E}}^2\left( {\bf{e}} \right)}}{{{{{E_s}} \mathord{\left/
 {\vphantom {{{E_s}} {\left( {4{N_0}} \right)}}} \right.
 \kern-\nulldelimiterspace} {\left( {4{N_0}} \right)}}}}} \right)}}} \right)
= \exp \left( { - \frac{{{E_s}}}{{8{N_0}}}d_{\rm{E}}^2\left( {\bf{e}} \right)} \right).
\end{align}
In particular, this approximation is reasonable since the value of $\exp \left( { - {{{{\left( {\sum\limits_{i = 1}^P {{\lambda _i}} } \right)}^2}} \mathord{\left/
 {\vphantom {{{{\left( {\sum\limits_{i = 1}^P {{\lambda _i}} } \right)}^2}} {\left( {2\sum\limits_{i = 1}^P {\lambda _i^2} } \right)}}} \right.
 \kern-\nulldelimiterspace} {\left( {2\sum\limits_{i = 1}^P {\lambda _i^2} } \right)}}} \right)$ changes slowly with the increase of ${2\sum\limits_{i = 1}^P {\lambda _{i}^2} }$, owing to the slow decay property of the corresponding function \cite{Yuan2003performance,vucetic2003space}.
On the other hand, the justification of the SNR assumption of (\ref{New_PEP_Large_P_derivation1}) is necessary. According to Lemma 4, we have
\begin{equation}
\frac{{P\sum\nolimits_{i = 1}^P {{\lambda _i}} }}{{\sum\nolimits_{i = 1}^P {\lambda _i^2} }} \le \frac{{{P^2}d_{\rm{E}}^2\left( {\bf{e}} \right)}}{{P{{\left( {d_{\rm{E}}^2\left( {\bf{e}} \right)} \right)}^2}}} = \frac{P}{{d_{\rm{E}}^2\left( {\bf{e}} \right)}}.
\end{equation}
where the equality only holds when the eigenvalues of ${\bf{\Omega }}\left( {\bf{e}} \right)$ share the same value, e.g., ${\bf{\Omega }}\left( {\bf{e}} \right)=\textrm{diag}\left\{ {d_{\rm{E}}^2{\left( {\bf{e}} \right)}, \ldots ,d_{\rm{E}}^2}{\left( {\bf{e}} \right)} \right\}$.
Therefore, we can see that the term $\frac{P{\sum\nolimits_{i = 1}^P {{\lambda _i}} }}{{\sum\nolimits_{i = 1}^P {\lambda _i^2} }}$ is upper-bounded by $\frac{P}{{d_{\rm{E}}^2\left( {\bf{e}} \right)}}$.
Hence, the assumption of SNR of (\ref{New_PEP_Large_P_derivation1}) can be further restricted as $\frac{{{E_s}}}{{4{N_0}}} \ge \frac{P}{{d_{\rm{E}}^2\left( {\bf{e}} \right)}}$.
This completes the proof of Theorem 3.

\bibliographystyle{IEEEtran}
\bibliography{OTFS_references}
\end{document}